\documentclass[aps,prev,twocolumn,preprintnumbers,floatfix,nofootinbib]{revtex4-1}
\pdfoutput=1
\usepackage{amssymb,graphicx}
\usepackage{epstopdf}
\usepackage{mathrsfs}
\usepackage{amssymb}
\usepackage{verbatim}
\usepackage{xcolor}
\usepackage{multirow}
\usepackage{amsmath}
\usepackage{braket}
\usepackage{subfig}
\usepackage{slashed} 
\usepackage{float}
\usepackage[normalem]{ulem}
\usepackage{sidecap}
\usepackage{hyperref}
\usepackage{cancel}
\usepackage{enumerate}
\hypersetup{pdfstartview=FitV,colorlinks=true,linkcolor=blue,citecolor=red,filecolor=black,urlcolor=blue}

\usepackage{tikz}
\usetikzlibrary{decorations.pathmorphing,arrows.meta,calc}

\tikzset{
  inflatonLine/.style={
    thick,
    dashed,
    draw=blue!80!black,
  },
  inflatonArrow/.style={
    thick,
    draw=blue!80!black,
    -{Latex[length=4pt,width=6pt]}, % larger arrows
  },
  graviton/.style={
    thick,
    draw=black,
    decorate,
    decoration={coil, aspect=0.7, segment length=4pt, amplitude=2pt},
  },
  spinTwoLine/.style={
    thick,
    draw=blue!80!black,
    double distance=0.8pt,
    decorate,
    decoration={snake, amplitude=1.4pt, segment length=5pt},
  },
  every node/.style={font=\small}
}

\usepackage[T1]{fontenc}
\usepackage[utf8]{inputenc}

\def\stacksymbols #1#2#3#4{\def\theguybelow{#2}
    \def\vp{\lower#3pt}
    \def\sp{\baselineskip0pt\lineskip#4pt}
    \mathrel{\mathpalette\intermediary#1}}

\def\intermediary#1#2{\vp\vbox{\sp
     \everycr={}\tabskip0pt
     \halign{$\mathsurround0pt#1\hfil##\hfil$\crcr#2\crcr
              \theguybelow\crcr}}}

\newcommand{\lslashslash}{%

  \raisebox{0.8ex}{%
    \scalebox{.7}{%
      \rotatebox[origin=c]{18}{$-$}%
    }%
  }%
}
\newcommand{\lslash}{%
  {%
   \vphantom{d}%
   \ooalign{\kern-.1em\smash{\lslashslash}\hidewidth\cr${\rm l}$\cr}%
   \kern.05em
  }%
}

\newcommand{\beq}{\begin{equation}}
\newcommand{\eeq}{\end{equation}}
\newcommand{\bea}{\begin{eqnarray}}
\newcommand{\eea}{\end{eqnarray}}

\newcommand{\gsim}{\lower.7ex\hbox{$\;\stackrel{\textstyle>}{\sim}\;$}}
\newcommand{\lsim}{\lower.7ex\hbox{$\;\stackrel{\textstyle<}{\sim}\;$}}

\def\be{\begin{equation}}
\def\ee{\end{equation}}
\def\bea{\begin{eqnarray}}
\def\eea{\end{eqnarray}}

\def\h{\eta}

\def\sp{\;\;\;,\;\;\;}

\def\lsim{\raise0.3ex\hbox{$\;<$\kern-0.75em\raise-1.1ex\hbox{$\sim\;$}}}
\def\gsim{\raise0.3ex\hbox{$\;>$\kern-0.75em\raise-1.1ex\hbox{$\sim\;$}}}

\def\inbar{\,\vrule height1.5ex width.4pt depth0pt}

\def\IC{\relax\hbox{$\inbar\kern-.3em{\rm C}$}}
\def\IQ{\relax\hbox{$\inbar\kern-.3em{\rm Q}$}}
\def\IR{\relax{\rm I\kern-.18em R}}
 \font\cmss=cmss10 \font\cmsss=cmss10 at 7pt
\def\IZ{\relax\ifmmode\mathchoice
 {\hbox{\cmss Z\kern-.4em Z}}{\hbox{\cmss Z\kern-.4em Z}}
 {\lower.9pt\hbox{\cmsss Z\kern-.4em Z}}
 {\lower1.2pt\hbox{\cmsss Z\kern-.4em Z}}\else{\cmss Z\kern-.4em Z}\fi}

\def\comment#1{}
\def\to{\rightarrow}

\def\u1x{U(1)_X}
\newcommand{\nc}{\newcommand}
\nc{\LL}{L}
\nc{\vv}{\tilde{v}}
\nc{\ccdot}{\!\cdot\!}
\nc{\gsm}{G_{SM}}
\nc{\vfive}{\mathbf{5}\oplus\mathbf{\overline{5}}}
\nc{\vten}{\mathbf{10}\oplus\mathbf{\overline{10}}}
\nc{\zhol}{Z^{\rm hol}}
\nc{\xfb}{\,{\rm fb}}

\begin{document}

%\wideabs{
%\begin{flushright}
%
%
%\end{flushright}

\vspace*{1mm}

\title{Gravitational Production of Massive Spin-2 Particles During Reheating}

\author{Sarunas Verner$^{a}$}
\email{verner@uchicago.edu}

\vspace{0.1cm}

\affiliation{
${}^a$ Kavli Institute for Cosmological Physics, 
University of Chicago, 5640 South Ellis Ave., Chicago, IL 60637, USA
 }

\begin{abstract} 
We study the minimal gravitational portal for a massive spin-2 dark matter candidate $X_{\mu\nu}$ produced during perturbative reheating. The dark sector couples to the visible sector only via gravity, and we analyze two unavoidable channels: (i) inflaton condensate annihilation, $\phi+\phi \to X+X$, and (ii) thermal scatterings, ${\rm SM}+{\rm SM}\to X+X$, both mediated by graviton exchange. Working in the Fierz-Pauli framework for a free massive spin-2 field of mass $m_2$, we derive the graviton-mediated amplitudes and perform a full helicity decomposition of the final state. The relic abundance is obtained analytically in terms of $m_2$ and the reheating temperature $T_{\rm RH}$. In the light mass regime $m_2 \ll m_\phi$ (with $m_\phi$ the inflaton mass during oscillations), production is overwhelmingly dominated by the longitudinal (helicity-0) mode: the $2\to2$ cross section is parametrically enhanced, scaling as $\sim (m_\phi/m_2)^4$, and yields efficient dark matter production despite purely gravitational couplings. Compared to lower-spin cases (spin-$0$, $1/2$, $1$, and $3/2$), massive spin-$2$ production is substantially more efficient for the same reheating history. Over most of the parameter space the inflaton condensate channel dominates the yield, while the thermal contribution is negligible. Avoiding overproduction typically requires either a relatively low $T_{\rm RH}$ or a spin-$2$ mass near threshold, $m_2 \lesssim m_\phi$. This places the spin-$2$ portal on similar footing to other higher spins in reheating scenarios, while emphasizing the central role of the helicity-$0$ mode and the reheating history in setting the dark matter density.
\end{abstract}

\maketitle

%%%%%%%%%%%%%%%%%%%%%%%%%%%%%%%%%%%%%%%%%%%%%%%%%%%%%%%%%%%%%%%%%%%%%%

%%%%%%%%%%%%%%%%%%%%%%%%%%%%%%%%%%%%%%%%%%%%%%%%%%%%%%%%%%%%%%%%%%%%%%
\section{Introduction}
\label{sec:intro}
%%%%%%%%%%%%%%%%%%%%%%%%%%%%%%%%%%%%%%%%%%%%%%%%%%%%%%%%%%%%%%%%%%%%%%
The identity and origin of dark matter (DM) remains unknown despite decades of increasingly sensitive searches~\cite{Zwicky:1933gu,Mambrini:2021cwd}. In particular, leading liquid xenon experiments have pushed spin-independent WIMP-nucleon limits to unprecedented levels, with recent analyses from LZ, XENONnT, and PandaX-4T reporting no statistically significant excess above backgrounds across the canonical GeV-TeV mass range~\cite{LZ:2022lsv, XENON:2025vwd, PandaX:2024qfu, LUX:2016ggv, Arcadi:2024ukq}. This sustained absence of discovery strengthens the case for DM candidates that are either very heavy, very light, very secluded, or produced through feeble interactions that are difficult to probe directly. Among the most model-independent possibilities are scenarios in which the dark sector is connected to the visible sector only through gravity, so that production is an inevitable consequence of the post-inflationary history rather than a model building choice.

Inflationary cosmology sharply motivates this perspective: once accelerated expansion ends, the Universe must be repopulated with a hot plasma~\cite{Dolgov:1982th, Abbott:1982hn, Nanopoulos:1983up}, with a reheating temperature at least above the few-MeV scale required for successful BBN and (in many baryogenesis mechanisms) potentially far higher. In perturbative reheating, the inflaton oscillates about the minimum of its potential and gradually transfers energy into relativistic species. Even when the inflaton has a direct decay channel to visible degrees of freedom, the thermal history is not characterized by a single temperature. Rather, the plasma typically reaches a maximum temperature $T_{\max}$ prior to the onset of radiation domination at $T_{\rm RH}$, and DM production can be parametrically sensitive to both scales~\cite{Giudice:2000ex,Chung:1998rq, Garcia:2017tuj, GKMO1}. This sensitivity becomes especially pronounced when the inflaton potential near its minimum deviates from purely quadratic behavior, or when DM is produced out of equilibrium (e.g.\ through freeze-in or ultraviolet-dominated processes), for which $T_{\max}$ can control the final yield~\cite{GKMO2,Bernal:2020gzm}.

Gravitational production provides a minimal ``floor'' for DM abundance during reheating.  Even in the absence of any non-gravitational portal, $2\to2$ processes mediated by the massless graviton unavoidably occur both within the thermal bath and directly from the coherently oscillating inflaton condensate. A systematic treatment of these channels, including production from inflaton scattering, production from the thermal bath, and even gravitational production of radiation itself, was developed in the gravitational portal framework of Refs.~\cite{Mambrini:2021zpp, Barman:2021ugy, Clery:2021bwz, Clery:2022wib}. Related analyses have emphasized that gravitational interactions can be relevant not only for the DM abundance but also for correlated phenomena during reheating such as leptogenesis and the emergence of a sufficiently hot plasma~\cite{Co:2022bgh, Barman:2022qgt}. More recently, complementary approaches have highlighted how the interplay of inflaton-mediated and gravity-mediated channels depends on the inflationary setup and the detailed reheating dynamics~\cite{Bernal:2024ykj,Henrich:2024rux}.

The idea that the post-inflationary Universe can populate a dark sector through
\emph{graviton-mediated} scattering has been developed systematically for lower-spin states, providing a useful baseline for assessing spin dependence during reheating~\cite{Mambrini:2021zpp, Bernal:2021kaj, Barman:2021ugy, Clery:2021bwz, Haque:2022kez, Clery:2022wib, Co:2022bgh, Barman:2022qgt, Kaneta:2023uwi, Ema:2015dka, Ema:2016hlw, Ema:2018ucl, Graham:2015rva, Garny:2015sjg, Garny:2017kha, Tang:2017hvq, Ema:2019yrd, Chianese:2020yjo, Chianese:2020khl, Ahmed:2020fhc, Kolb:2020fwh, Redi:2020ffc, Ling:2021zlj, Ahmed:2021fvt, Haque:2021mab, Aoki:2022dzd, Ahmed:2022tfm, Haque:2023yra, Kaneta:2022gug, Kolb:2023dzp, Kolb:2023ydq, Kaneta:2023uwi, Garcia:2023obw, Kaneta:2023kfv, Garcia:2022vwm, Garcia:2023awt, Garcia:2023qab, Zhang:2023xcd, Ozsoy:2023gnl, Cembranos:2023qph}. For spin-$0$ and spin-$\frac12$ dark matter, graviton exchange generates an irreducible abundance from both inflaton condensate scattering in the oscillatory era and from
scatterings in the thermal bath, with yields that can be sensitive to the full reheating history via $T_{\max}$ and $T_{\rm RH}$~\cite{Ema:2018ucl, Mambrini:2021zpp, Barman:2021ugy, Clery:2021bwz, Clery:2022wib, Ling:2021zlj}. For spin-1, the situation is more nuanced because of conformal effects in the massless limit, but for \emph{massive} vectors both inflationary gravitational production and reheating-era scattering sources have been analyzed, including the role of single-graviton exchange during reheating \cite{Garcia:2023obw}. For spin-$\frac32$, graviton-mediated production during reheating has likewise been shown to exhibit a characteristic hierarchy between condensate-sourced and bath-sourced channels in broad regions of parameter space \cite{Kaneta:2023uwi}. Together, these studies emphasize that the helicity content of the final state controls both the dominant production channel and the parametric dependence on $T_{\max}$ versus $T_{\rm RH}$.

Two broad production mechanisms are then immediately relevant for a \emph{massive spin-2} final state. On the one hand, massive spin-2 quanta can be generated by cosmological gravitational particle production (CGPP) of vacuum fluctuations across inflation and the transition to the hot Big Bang~\cite{Kolb:2023dzp}. On the other hand, once the inflaton begins oscillating and a plasma is formed, spin-2 states can be populated by ordinary $2\to2$ scatterings mediated by \emph{single exchange of the massless graviton}. The latter contribution is present in the complete absence of any non-gravitational portal and is fixed, up to the reheating background, by the universal coupling of the graviton to stress-energy tensor. In what follows, we adopt this minimal setup: the \emph{produced} dark matter is a massive Fierz-Pauli spin-2 field,\footnote{For studies considering massive spin-2 dark matter, see Refs.~\cite{Aoki:2016zgp, Babichev:2016hir, Gorji:2023cmz, Manita:2022tkl, Kolb:2023dzp}} while the \emph{mediator} is the massless spin-2 graviton. 

A comprehensive cosmological analysis of gravitational particle production of massive spin-2 quanta across \emph{inflation and reheating} was carried out in Ref.~\cite{Kolb:2023dzp}. That study solved the full system of mode equations (including the helicity-0 sector) for representative inflationary potentials, computed the resulting spectra and abundances, and identified conditions for ghost or gradient instabilities, deriving a generalized Higuchi bound in FRW backgrounds~\cite{Higuchi:1986py}. The conclusion is that cosmological gravitational particle production can yield the observed DM abundance in viable regions of parameter space, while simultaneously imposing non-trivial consistency conditions on the effective description. Our goal here is complementary. We develop an analytic and semi-analytic treatment of \emph{post-inflationary} production during perturbative reheating using a scattering picture, placing massive spin-2 DM on the same footing as the gravitational-portal literature for scalars and fermions~\cite{Mambrini:2021zpp, Clery:2021bwz}. This has two advantages. First, it isolates the dependence of the yield on $T_{\rm RH}$ and the inflaton equation of state in a transparent way. Second, it clarifies how the spin-2 polarization structure reshapes the condensate-scattering and bath-scattering rates relative to lower-spin final states, and which helicities dominate in the relevant regimes.

We consider a massive spin-2 field $X_{\mu\nu}$ described at the free level by the Fierz-Pauli Lagrangian~\cite{Fierz:1939zz, Fierz:1939ix} and coupled universally to the stress-energy tensor. Within this setup, we compute the minimal production of $X_{\mu\nu}$ through
(i) inflaton condensate scattering during the oscillatory phase and (ii) graviton-mediated scattering in the thermal bath after inflaton decays populate the plasma. We treat reheating perturbatively, allow for generic reheating histories (parameterized by $T_{\rm RH}$), and keep the full mass dependence relevant for both production mechanisms. We then determine the relic abundance and map the regions in which massive spin-2 particles can constitute all of DM while remaining consistent with basic cosmological requirements. 

The remainder of this paper is organized as follows. In Sec.~\ref{sec:gravprodspin2} we summarize the Fierz-Pauli description of a free massive spin-2 field and fix our conventions. We also derive the graviton couplings relevant for $2\to2$ production and set up the polarization sums and helicity structure needed for massive spin-2 final states. In Sec.~\ref{sec:abundance} we compute production from the inflaton condensate during reheating and obtain the corresponding yield and relic density in terms of $m_2$, $T_{\rm RH}$, and the inflaton mass $m_{\phi}$. In Sec.~\ref{sec:spin2dmthermal} we evaluate the graviton-mediated thermal production from the SM bath. We conclude in Sec.~\ref{sec:summary}. Technical details are collected in the appendices.

%%%%%%%%%%%%%%%%%%%%%%%%%%%%%%%%%%%%%%%%%%%%%%%%%%%%%%%%%%%%%%%%
\section{Gravitational production of a massive spin-2 field}
\label{sec:gravprodspin2}
%%%%%%%%%%%%%%%%%%%%%%%%%%%%%%%%%%%%%%%%%%%%%%%%%%%%%%%%%%%%%%%%
We now extend the gravitational production framework to a neutral, massive spin-2 field that interacts only through gravity. Throughout this work, we denote the spin-2 field by $X_{\mu\nu}$ and assume that it is stable on cosmological time scales, so that it can contribute to the dark matter density. Its interactions with the inflaton condensate and with the thermal Standard Model (SM) bath are mediated solely by the canonical metric perturbation $h_{\mu\nu}$, defined via the expansion of the space-time metric around Minkowski space,
\begin{equation}
  g_{\mu\nu} \;=\; \eta_{\mu\nu} + \frac{2}{M_P} h_{\mu\nu}\,,
  \label{eq:metric_expansion}
\end{equation}
where $M_P$ is the reduced Planck mass. During reheating the Hubble rate is much smaller than the characteristic momentum transfer in the scattering processes and in the inflaton oscillations, so that the description in Eq.~\eqref{eq:metric_expansion} is sufficient for computing the
relevant amplitudes. In this regime gravitational interactions are universal and couple the inflaton, the SM plasma, and the massive spin-2 field $X_{\mu\nu}$
through their energy-momentum tensors. A stable $X_{\mu\nu}$ can therefore be populated either from the oscillating inflaton condensate or from the SM thermal bath and act as a viable dark matter candidate.

\begin{figure}[h!]
\centering
\begin{tikzpicture}[baseline=(current bounding box.center)]
  % coordinates (slightly larger angles)
  \coordinate (L1) at (-3,  1.6);
  \coordinate (L2) at (-3, -1.6);
  \coordinate (VL) at (-0.9,  0);
  \coordinate (VR) at ( 0.9,  0);
  \coordinate (R1) at ( 3,  1.6);
  \coordinate (R2) at ( 3, -1.6);

  % inflaton / SM: dashed + arrow in the middle
  \draw[inflatonLine] (L1) -- (VL);
  \draw[inflatonArrow] ($(L1)!0.45!(VL)$) -- ($(L1)!0.55!(VL)$);

  \draw[inflatonLine] (L2) -- (VL);
  \draw[inflatonArrow] ($(L2)!0.45!(VL)$) -- ($(L2)!0.55!(VL)$);

  % spin-2 external legs: double wiggly, no arrows
  \draw[spinTwoLine] (VR) -- (R1);
  \draw[spinTwoLine] (VR) -- (R2);

  % graviton propagator (black)
  \draw[graviton] (VL) -- (VR);

  % labels
  \node[left=2pt]  at (L1) {$\phi/{\rm SM}$};
  \node[left=2pt]  at (L2) {$\phi/{\rm SM}$};

  \node[right=2pt] at (R1) {$X_{\mu\nu}$};
  \node[right=2pt] at (R2) {$X_{\mu\nu}$};

  \node[above=2pt] at ($(VL)!0.5!(VR)$) {$h_{\mu\nu}$};

  \node[above=4pt] at (VL)
    {$\displaystyle \frac{T^{\mu\nu}_{\phi/{\rm SM}}}{M_P}$};
  \node[above=4pt] at (VR)
    {$\displaystyle \frac{T^{\mu\nu}_{(2)}}{M_P}$};
\end{tikzpicture}

\caption{Feynman diagram for the gravitational production of massive
spin-2 particles from inflaton or SM scattering.}
\label{fig:spin2portal}
\end{figure}
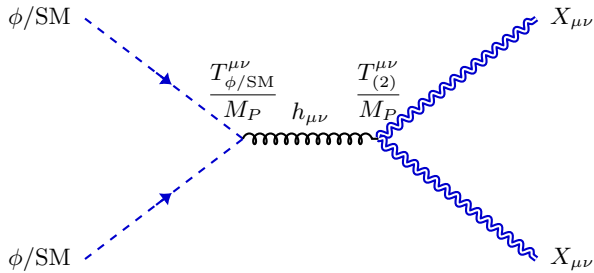

The corresponding gravitational portal is illustrated in
Fig.~\ref{fig:spin2portal}, which shows the production of massive
spin-2 particles from inflaton or SM scattering via exchange of the
graviton $h_{\mu\nu}$.

%%%%%%%%%%%%%%%%%%%%%%%%%%%%%%%%%%%%%%%%%%%%%%%%%%%%%%%%%%%%%%%%
\subsection{Massive spin-2 field: Fierz-Pauli theory}
\label{sec:FP}
%%%%%%%%%%%%%%%%%%%%%%%%%%%%%%%%%%%%%%%%%%%%%%%%%%%%%%%%%%%%%%%%
A free massive spin-2 field in flat space is described by the
Fierz-Pauli theory~\cite{Fierz:1939zz, Fierz:1939ix}. We denote the field by
$X_{\mu\nu}=X_{\nu\mu}$ and work in Minkowski space with metric
$\eta_{\mu\nu}=\mathrm{diag}(+1,-1,-1,-1)$ and trace
$X\equiv \eta^{\mu\nu}X_{\mu\nu}$.  The Fierz-Pauli Lagrangian density is
\begin{equation}
\begin{aligned}
  &\mathcal{L}_{\rm FP}
  \;=\;
  -\frac12\,\partial_\lambda X_{\mu\nu}\,\partial^\lambda X^{\mu\nu}
  + \partial_\mu X^{\mu\nu}\,\partial^\lambda X_{\lambda\nu} \\
  &- \partial_\mu X^{\mu\nu}\,\partial_\nu X
  + \frac12\,\partial_\lambda X\,\partial^\lambda X
  - \frac12 m_2^2 \bigl(X_{\mu\nu}X^{\mu\nu}-X^2\bigr)\,,
  \label{eq:FP_Lagrangian}
\end{aligned} 
\end{equation}
where $m_2$ is the spin-2 mass. Up to total derivatives and an
overall sign convention, this is the standard Fierz-Pauli Lagrangian used in the massive-gravity and spin-2 literature (see, e.g.,
Refs.~\cite{Hinterbichler:2011tt,deRham:2014zqa,Folkerts:2013mra, Koenigstein:2015asa, Farolfi:2025knq}).

Varying Eq.~\eqref{eq:FP_Lagrangian} with respect to $X^{\mu\nu}$ gives
the equations of motion
\begin{equation}
\begin{aligned}
  \Box X_{\mu\nu}
  &- 2\,\partial_{(\mu}\partial^\rho X_{\nu)\rho}
  + \partial_\mu\partial_\nu X
  + \eta_{\mu\nu}\,\partial^\rho\partial^\sigma X_{\rho\sigma}
  \\
  &- \eta_{\mu\nu}\,\Box X
  \;=\;
  m_2^2\bigl(X_{\mu\nu}-\eta_{\mu\nu}X\bigr)\,,
  \label{eq:FP_EOM_full}
\end{aligned}
\end{equation}
where $\Box\equiv\partial_\rho\partial^\rho$ and parentheses denote
symmetrization,
$A_{(\mu}B_{\nu)}\equiv (A_\mu B_\nu + A_\nu B_\mu)/2$.
Taking the divergence of Eq.~\eqref{eq:FP_EOM_full} and using the
commutativity of partial derivatives yields
\begin{equation}
  \partial^\mu X_{\mu\nu} - \partial_\nu X = 0\,,
  \label{eq:FP_constraint_div}
\end{equation}
while taking the trace and combining with
Eq.~\eqref{eq:FP_constraint_div} gives
\begin{equation}
  X = 0 \,.
  \label{eq:FP_constraint_trace}
\end{equation}
Equations~\eqref{eq:FP_EOM_full}-\eqref{eq:FP_constraint_trace} are
equivalent to the simpler \emph{Fierz-Pauli system}
\begin{align}
  (\Box - m_2^2)\,X_{\mu\nu} &= 0\,,
  \label{eq:FP_KG}\\[2pt]
  \partial^\mu X_{\mu\nu} &= 0\,,
  \label{eq:FP_transverse}\\[2pt]
  X &= 0\,,
  \label{eq:FP_traceless}
\end{align}
which shows that $X_{\mu\nu}$ is transverse and traceless and satisfies
a Klein-Gordon equation, propagating the expected five physical polarizations of a massive spin-$2$ field.

The mass term in Eq.~\eqref{eq:FP_Lagrangian},
\begin{equation}
  \mathcal{L}_{\rm mass}
  \;=\;
  -\frac12 m_2^2\bigl(X_{\mu\nu}X^{\mu\nu}-X^2\bigr)\,,
\end{equation}
is uniquely fixed by the requirement that no additional ghostlike
scalar mode propagates: any other Lorentz-invariant quadratic
combination of $X_{\mu\nu}$ and $X$ leads to a sixth degree of freedom
with wrong-sign kinetic energy~\cite{Fierz:1939ix,deRham:2014zqa,Hinterbichler:2011tt}.
In the massless limit $m_2\to 0$, the Lagrangian
\eqref{eq:FP_Lagrangian} reduces to the linearized Einstein-Hilbert
action and becomes invariant under the gauge symmetry $\delta X_{\mu\nu} = \partial_\mu\xi_\nu + \partial_\nu\xi_\mu$, which corresponds to linearized diffeomorphisms.

For completeness we note that consistent extensions of the FP theory
to curved backgrounds and to non-linear massive gravity are discussed
in detail in Refs.~\cite{Hinterbichler:2011tt,deRham:2014zqa,Folkerts:2013mra, Koenigstein:2015asa, Farolfi:2025knq} and references therein. In this work we only require the flat-space theory summarized above.

%%%%%%%%%%%%%%%%%%%%%%%%%%%%%%%%%%%%%%%%%%%%%%%%%%%%%%%%%%%%%%%%
\subsection{Gravitational couplings}
\label{sec:couplings_spin2}
%%%%%%%%%%%%%%%%%%%%%%%%%%%%%%%%%%%%%%%%%%%%%%%%%%%%%%%%%%%%%%%%    
To linear order in $\h_{\mu\nu}$, the coupling to matter and to the
spin-2 field is encoded in the universal interaction
\begin{equation}
  \mathcal{L}_{\rm int}
  \;=\;
  -\frac{1}{M_P}\,
  h_{\mu\nu}
  \left(
    T^{\mu\nu}_{\rm SM}
    + T^{\mu\nu}_\phi
    + T^{\mu\nu}_{2}
  \right)\!,
  \label{Eq:lagrangian_spin2}
\end{equation}
where $\phi$ is the inflaton, ``SM'' stands for the Standard Model
sector, and $T^{\mu\nu}_{2}$ is the energy-momentum tensor of the
massive spin-2 field $X_{\mu\nu}$.

For the inflaton and SM fields we adopt the standard canonical
energy-momentum tensors for spins $0$, $1/2$, and $1$,
\begin{align}
  T^{\mu\nu}_{0}
  &=\;
  \partial^\mu S\,\partial^\nu S
  - \eta^{\mu\nu}
  \left[
    \frac12 \partial^\alpha S\,\partial_\alpha S - V(S)
  \right] ,
  \label{Eq:tensors_spin2}
  \\[4pt]
  T^{\mu\nu}_{1/2}
  &=\;
  \frac{i}{4}
  \left[
    \bar\chi \gamma^\mu \overset{\leftrightarrow}{\partial^\nu} \chi
    + \bar\chi \gamma^\nu \overset{\leftrightarrow}{\partial^\mu} \chi
  \right]
  \nonumber\\[-2pt]
  &\quad
  - \eta^{\mu\nu}
  \left[
    \frac{i}{2}
    \bar\chi \gamma^\alpha \overset{\leftrightarrow}{\partial_\alpha} \chi
    - m_\chi \bar\chi \chi
  \right] ,
  \label{Eq:tensorf_spin2}
  \\[4pt]
  T^{\mu\nu}_{1}
  &=\;
  \frac12
  \Big[
    F^\mu{}_{\alpha} F^{\nu\alpha}
    + F^\nu{}_{\alpha} F^{\mu\alpha}
    - \frac12 \eta^{\mu\nu} F^{\alpha\beta} F_{\alpha\beta}
  \Big] ,
  \label{Eq:tensorv_spin2}
\end{align}
where $S = \phi, H$ denotes either the inflaton or the (real) Higgs
scalar,\footnote{In the numerical calculations, we treat the Higgs field
as four real scalars, corresponding to the four degrees of freedom of
the SM Higgs doublet.} $V(S)$ is the scalar potential,
$A \overset{\leftrightarrow}{\partial_\mu} B \equiv
 A\,\partial_\mu B - (\partial_\mu A)\,B$, and $F_{\mu\nu} = \partial_\mu A_\nu - \partial_\nu A_\mu$ is the vector field strength.

For the massive spin-2 field, we start from the covariant Fierz-Pauli
Lagrangian~(\ref{eq:FP_Lagrangian}) and define its energy-momentum tensor in the
usual way,
\begin{equation}
  T^{\mu\nu}_{2}
  \;\equiv\;
  -\frac{2}{\sqrt{-g}}\,
  \frac{\delta \mathcal{L}_{\rm FP}}{\delta g_{\mu\nu}}\,.
  \label{Eq:T_spin2_def}
\end{equation}
For physical (on-shell) Fierz-Pauli solutions in Minkowski space,
$\partial^\mu X_{\mu\nu}=0$ and $X^\mu{}_\mu = 0$, this reduces to
the compact form~\cite{Rivers:1964nfl, Koenigstein:2015asa}
\begin{equation}
\begin{aligned}
  T^{\mu\nu}_{2}
  \;=\;
  &\partial^\mu X_{\alpha\beta}\,\partial^\nu X^{\alpha\beta}
 \\
 - &\eta^{\mu\nu}\,
    \frac12
    \Big[
      \partial_\rho X_{\alpha\beta}\,\partial^\rho X^{\alpha\beta}
      - m_2^2\,X_{\alpha\beta} X^{\alpha\beta}
    \Big] \, .
  \label{Eq:T_spin2_on_shell}
\end{aligned} 
\end{equation}
On shell, Eq.~\eqref{Eq:T_spin2_on_shell} is conserved and symmetric
and carries the five propagating polarizations of the massive spin-2
field.

The tree-level amplitudes relevant for the production of massive
spin-2 particles through gravitational scatterings,
\begin{equation}
  \phi/{\rm SM}^i(p_1) + \phi/{\rm SM}^i(p_2)
  \;\rightarrow\;
  X(p_3) + X(p_4)\,,
  \label{Eq:process_spin2}
\end{equation}
are mediated by a single graviton exchange with momentum $k = p_1 + p_2$.
It is convenient to factorize the amplitude in terms of ``partial'' tensors contracted with the graviton propagator,
\begin{equation}
  \mathcal{M}^{i\,2}
  \;\propto\;
  M^{2}_{\mu\nu}\,
  \Pi^{\mu\nu\rho\sigma}(k)\,
  M^{i}_{\rho\sigma}\,,
  \label{Eq:scamp_spin2}
\end{equation}
with $i=0, 1/2, 1$ indicating the spin of the initial-state
particles. 

The graviton propagator for the canonical field $h_{\mu\nu}$ in
de~Donder gauge is~\cite{Giudice:1998ck}
\begin{equation}
  \Pi^{\mu\nu\rho\sigma}(k)
  \;=\;
  \frac{
    \eta^{\mu\rho}\eta^{\nu\sigma}
    + \eta^{\mu\sigma}\eta^{\nu\rho}
    - \eta^{\mu\nu}\eta^{\rho\sigma}
  }{2\,k^2}\,.
  \label{Eq:graviton_prop}
\end{equation}
The partial tensors $M^i_{\mu\nu}$ associated with the initial states
are defined as matrix elements of the corresponding energy-momentum tensors and can be written as~\cite{Clery:2021bwz}
\begin{align}
  M^{0}_{\mu\nu}
  &=\;
  \frac{1}{2}
  \Big[
    p_{1\mu} p_{2\nu}
    + p_{1\nu} p_{2\mu}
    - \eta_{\mu\nu}\, p_1\!\cdot\! p_2
    - \eta_{\mu\nu}\, V''(S)
  \Big] ,
  \label{Eq:partamp_scalar_spin2}
  \\[4pt]
  M^{1/2}_{\mu\nu}
  &=\;
  \frac{1}{4}\,
  \bar v(p_2)
  \Big[
    \gamma_\mu (p_1 - p_2)_\nu
    + \gamma_\nu (p_1 - p_2)_\mu
  \Big] u(p_1)\,,
  \label{Eq:partamp_fermion_spin2}
  \\[4pt]
  M^{1}_{\mu\nu}
  &=\;
  \frac12 \bigg[
    \epsilon_2^{*}\!\cdot\!\epsilon_1
      \big(p_{1\mu} p_{2\nu} + p_{1\nu} p_{2\mu}\big)
    - \epsilon_2^{*}\!\cdot\!p_1
      \big(p_{2\mu} \epsilon_{1\nu} + \epsilon_{1\mu} p_{2\nu}\big)
    \nonumber\\[-2pt]
  &\qquad
    - \epsilon_1\!\cdot\!p_2
      \big(p_{1\nu} \epsilon^{*}_{2\mu} + p_{1\mu} \epsilon^{*}_{2\nu}\big)
    + p_1\!\cdot\!p_2
      \big(\epsilon_{1\mu} \epsilon^{*}_{2\nu}
         + \epsilon_{1\nu} \epsilon^{*}_{2\mu}\big)
    \nonumber\\[2pt]
  &\qquad
    + \eta_{\mu\nu}
      \big(
        \epsilon_2^{*}\!\cdot\!p_1\, \epsilon_1\!\cdot\!p_2
        - p_1\!\cdot\!p_2\, \epsilon_2^{*}\!\cdot\!\epsilon_1
      \big)
  \bigg] ,
  \label{Eq:partamp_vector_spin2}
\end{align}
where we neglect the masses of SM fermions and gauge bosons.

The massive spin-2 field $X_{\mu\nu}$ is expanded in creation and
annihilation operators as
\begin{equation}
\begin{aligned}
  &X_{\alpha\beta}(x) \; = \;
  \sum_{\lambda} \int\!\frac{d^3\mathbf p}{(2\pi)^3 2E_{\mathbf p}} \times \\
  &\left[
    \epsilon_{\alpha\beta}(p,\lambda) a_{p,\lambda} e^{-ip\cdot x}
    + \epsilon^*_{\alpha\beta}(p,\lambda) a^\dagger_{p,\lambda} e^{+ip\cdot x}
  \right],
 \end{aligned} 
\end{equation}
with polarization tensors satisfying
\begin{equation}
  p^\alpha \epsilon_{\alpha\beta}(p,\lambda)=0,\qquad
  \epsilon^\alpha{}_{\alpha}(p,\lambda)=0,\qquad
  \epsilon_{\alpha\beta}=\epsilon_{\beta\alpha},
\end{equation}
and normalized as
\begin{equation}
  \epsilon_{\alpha\beta}(p,\lambda)\,
  \epsilon^{*\,\alpha\beta}(p,\lambda') = \delta_{\lambda\lambda'}.
\end{equation}

Using the form~\eqref{Eq:T_spin2_on_shell} and the mode
expansion of $X_{\alpha\beta}$, one finds the compact form
\begin{equation}
\begin{aligned}
  M^{2}_{\mu\nu} &\; = \; \frac12
  \Big[
    p_{3\mu} p_{4\nu}
    + p_{3\nu} p_{4\mu}
    - \eta_{\mu\nu}\bigl(p_3\!\cdot\!p_4 + m_2^2\bigr)
  \Big]
   \\
 & \; \times \;
  \epsilon_{\alpha\beta}(p_3,\lambda_3)\,
  \epsilon^{\alpha\beta}(p_4,\lambda_4)\,,
  \label{Eq:M_spin2_partial}
\end{aligned}  
\end{equation}
where $\epsilon_{\alpha\beta}(p,\lambda)$ are the massive spin-2
polarization tensors defined above. The polarization sum entering the
squared amplitude can be expressed in terms of the standard massive
spin-2 projector. We will use this below to obtain the
polarization-averaged squared matrix element relevant for the
gravitational production of $X_{\mu\nu}$.

There are two distinct gravitational production mechanisms, both
represented by the diagram in Fig.~\ref{fig:spin2portal} with the final
state of the massive spin-2 field:
\begin{itemize}
  \item[(i)] Production from the inflaton condensate,
    $\phi + \phi \rightarrow X + X$.
    For a purely quadratic minimum, the inflaton behaves as a
    collection of non-relativistic quanta at rest, and the
    corresponding partial tensor is directly given by
    Eq.~\eqref{Eq:partamp_scalar_spin2} with $S = \phi$.
    For more general minima
    $V(\phi) \simeq \lambda \phi^k / M_P^{k-4}$ we instead use the
    zero-mode description of the oscillating condensate, discussed below.

  \item[(ii)] Production from the thermal plasma,
    ${\rm SM} + {\rm SM} \rightarrow X + X$,
    where the initial-state tensors $M^{i}_{\mu\nu}$ with
    $i=0,1/2,1$ are those in
    Eqs.~\eqref{Eq:partamp_scalar_spin2}-\eqref{Eq:partamp_vector_spin2}.
\end{itemize}
In both channels the spin-2 field is coupled only gravitationally, so
its production rate is completely determined by the Fierz-Pauli
dynamics and by the universal graviton exchange encoded in
Eq.~\eqref{Eq:scamp_spin2}. In the following, we keep the inflaton
potential as general as possible, restricting ourselves to models that satisfy the slow-roll constraints from \textit{Planck}~\cite{Planck:2018vyg} and related CMB measurements, and that admit a well-defined minimum allowing the
expansion $V(\phi) \simeq \lambda \phi^k/M_P^{\,k-4}$ around it.

%%%%%%%%%%%%%%%%%%%%%%%%%%%%%%%%%%%%%%%%%%%%%%%%%%%%%%%%%%%%%%%%
\subsection{Spin-2 polarization tensors and SVT decomposition}
\label{sec:svt}
%%%%%%%%%%%%%%%%%%%%%%%%%%%%%%%%%%%%%%%%%%%%%%%%%%%%%%%%%%%%%%%%
To resolve the helicity content of the massive spin-2 final state, we
construct its polarization tensors from the massive spin-1
polarizations.  For a momentum $p^\mu=(E,\mathbf p)$ with
$p^2 = m_2^2$ and metric $(+,-,-,-)$, we use the three spin-1
polarizations $\epsilon_r^\mu(p)$ with $r=\{+,0,-\}$,
\begin{equation}
  p_\mu \epsilon_r^\mu(p) = 0\,, \qquad
  \epsilon_r\!\cdot\!\epsilon_{r'}^*(p) = \delta_{rr'}\,,
\end{equation}
given explicitly by
\begin{align}
  \epsilon_{+}^\mu(p)
  &= \frac{1}{\sqrt{2}}
  \left(\!\begin{array}{c}
    0 \\
    -\cos\theta\cos\phi + i\sin\phi \\
    -\cos\theta\sin\phi - i\cos\phi \\
    \phantom{-}\sin\theta
  \end{array}\!\right), \\[4pt]
  \epsilon_{0}^\mu(p)
  &= \frac{1}{m_2}
  \left(\!\begin{array}{c}
    |\mathbf p| \\
    E\sin\theta\cos\phi \\
    E\sin\theta\sin\phi \\
    E\cos\theta
  \end{array}\!\right), \\[4pt]
  \epsilon_{-}^\mu(p)
  &= \frac{1}{\sqrt{2}}
  \left(\!\begin{array}{c}
    0 \\
    \cos\theta\cos\phi + i\sin\phi \\
    \cos\theta\sin\phi - i\cos\phi \\
    -\sin\theta
  \end{array}\!\right),
\end{align}
which correspond to helicities $\lambda=\pm 1,0$.

A massive spin-2 polarization tensor is the symmetric, traceless part
of the tensor product of two spin-1 polarizations. Using the
Clebsch-Gordan coefficients for $1\otimes 1\to 2$, we define the
five helicity tensors $\epsilon^{\mu\nu}_\lambda(p)$,
$\lambda=\pm2,\pm1,0$, as
\begin{align}
  \epsilon^{\mu\nu}_{+2}(p)
  &= \epsilon^\mu_{+}(p)\,\epsilon^\nu_{+}(p),\\[3pt]
  \epsilon^{\mu\nu}_{+1}(p)
  &= \frac{1}{\sqrt{2}}\Big(
      \epsilon^\mu_{+}(p)\,\epsilon^\nu_{0}(p)
      + \epsilon^\mu_{0}(p)\,\epsilon^\nu_{+}(p)
    \Big),\\[3pt]
  \epsilon^{\mu\nu}_{0}(p)
  &= \frac{1}{\sqrt{6}}\Big(
      \epsilon^\mu_{+}(p)\,\epsilon^\nu_{-}(p)
      + \epsilon^\mu_{-}(p)\,\epsilon^\nu_{+}(p)
      + 2\,\epsilon^\mu_{0}(p)\,\epsilon^\nu_{0}(p)
    \Big),\\[3pt]
  \epsilon^{\mu\nu}_{-1}(p)
  &= \frac{1}{\sqrt{2}}\Big(
      \epsilon^\mu_{-}(p)\,\epsilon^\nu_{0}(p)
      + \epsilon^\mu_{0}(p)\,\epsilon^\nu_{-}(p)
    \Big),\\[3pt]
  \epsilon^{\mu\nu}_{-2}(p)
  &= \epsilon^\mu_{-}(p)\,\epsilon^\nu_{-}(p).
\end{align}
These tensors are symmetric, transverse and traceless,
\begin{equation}
  \epsilon^{\mu\nu}_\lambda(p)
  = \epsilon^{\nu\mu}_\lambda(p),\qquad
  p_\mu \epsilon^{\mu\nu}_\lambda(p) = 0,\qquad
  \eta_{\mu\nu}\epsilon^{\mu\nu}_\lambda(p)=0,
\end{equation}
and orthonormal,
\begin{equation}
  \epsilon^{\mu\nu}_\lambda(p)\,
  \epsilon^{*}_{\mu\nu,\lambda'}(p) = \delta_{\lambda\lambda'}.
\end{equation}

To sum over the five physical polarizations of the final-state
spin-2 particles, it is convenient to introduce the standard massive
spin-2 polarization projector,
\begin{equation}
  \sum_{\lambda}
  \epsilon^{\lambda}_{\mu\nu}(p)\,
  \epsilon^{* \, \lambda}_{\rho\sigma}(p)
  \;=\;
  \mathcal P_{\mu\nu,\rho\sigma}(p)\,,
  \label{eq:spin2_projector_def}
\end{equation}
with
\begin{equation}
  \mathcal P_{\mu\nu,\rho\sigma}(p)
  \;\equiv\;
  \frac12\Big(
    G_{\mu\rho}G_{\nu\sigma}
    + G_{\mu\sigma}G_{\nu\rho}
  \Big)
  - \frac13\,G_{\mu\nu}G_{\rho\sigma}\,,
  \label{eq:spin2_projector}
\end{equation}
where
\begin{equation}
      G_{\mu\nu} \equiv \eta_{\mu\nu} - \frac{p_\mu p_\nu}{m_2^2}\,.
\end{equation}

For later use, it is convenient to rewrite this projector in a scalar-vector-tensor (SVT) form that separates the contributions of the five helicities. We also choose a fixed timelike reference vector $n^\mu$ with $n^2=1$ and $p\cdot n \neq 0$, and define the longitudinal (helicity-0) polarization vector of the massive spin-1 as
\begin{equation}
  e_L^\mu(p)
  \;=\;
  \frac{(p\!\cdot n)\,p^\mu - m_2^2\,n^\mu}
       {m_2\,\sqrt{(p\!\cdot n)^2 - m_2^2}}\,,
\end{equation}
which satisfies $ p_\mu e_L^\mu = 0$ and $e_L\!\cdot e_L = -1$ in our $(+,-,-,-)$ convention. The projector onto the two transverse
vector polarizations ($\lambda=\pm1$) is then
\begin{equation}
  P_T^{\mu\nu}(p)
  \;\equiv\;
  G^{\mu\nu}(p) - e_L^\mu(p)\,e_L^\nu(p) \, .
\end{equation}

In terms of $P_T^{\mu\nu}$ and $e_L^\mu$ the spin-2 SVT projectors
take the form
\begin{align}
  \mathcal P^{\mu\nu,\rho\sigma}_{T}(p)
  &= \frac12\!\left[
        P_T^{\mu\rho} P_T^{\nu\sigma}
      + P_T^{\mu\sigma} P_T^{\nu\rho}
      \right]
     - \frac12\,P_T^{\mu\nu} P_T^{\rho\sigma} \, ,
  \label{eq:P2_tensor}
  \\[4pt]
  \mathcal P^{\mu\nu,\rho\sigma}_{V}(p)
  &= \frac12\!\left[
      P_T^{\mu\rho} e_L^{\nu} e_L^{\sigma}
    + P_T^{\mu\sigma} e_L^{\nu} e_L^{\rho}
    + P_T^{\nu\rho} e_L^{\mu} e_L^{\sigma}
    + P_T^{\nu\sigma} e_L^{\mu} e_L^{\rho}
    \right] \, ,
  \label{eq:P2_vector}
  \\[4pt]
  \mathcal P^{\mu\nu,\rho\sigma}_{S}(p)
  &= \frac16\,
    \big(2 e_L^{\mu} e_L^{\nu} - P_T^{\mu\nu}\big)\,
    \big(2 e_L^{\rho} e_L^{\sigma} - P_T^{\rho\sigma}\big) \, ,
  \label{eq:P2_scalar}
\end{align}
corresponding respectively to the helicity sectors
$\lambda=\pm 2$ (tensor), $\lambda=\pm 1$ (vector) and $\lambda=0$
(scalar).  One can check explicitly that these projectors satisfy
\begin{equation}
  \mathcal P_{A}\, \mathcal P_{B}
  = \delta_{AB}\,\mathcal P_{A},
  \qquad
  \sum_{A=T,V,S} \mathcal P_{A}^{\mu\nu,\rho\sigma}(p)
  =
  \mathcal P^{\mu\nu,\rho\sigma}(p),
\end{equation}
so that they provide a complete and orthogonal decomposition of the
massive spin-2 projector into tensor, vector, and scalar components. 

The contributions of each helicity sector to the squared amplitude
follow by contracting these projectors with the partial tensor
$M^{(2)}_{\mu\nu}$,
\begin{equation}
\begin{aligned}
  \big|\mathcal M^{(2)}\big|^2_{S}
  &= M^{2}_{\mu\nu}\,
     \mathcal P^{\mu\nu,\rho\sigma}_{S}(p_3)\,
     M^{2\,*}_{\rho\sigma},\\[2pt]
  \big|\mathcal M^{(2)}\big|^2_{V}
  &= M^{2}_{\mu\nu}\,
     \mathcal P^{\mu\nu,\rho\sigma}_{V}(p_3)\,
     M^{2\,*}_{\rho\sigma},\\[2pt]
     \big|\mathcal M^{(2)}\big|^2_{T}
  &= M^{2}_{\mu\nu}\,
     \mathcal P^{\mu\nu,\rho\sigma}_{T}(p_3)\,
     M^{2\,*}_{\rho\sigma} ,
\end{aligned}
\end{equation}
and
\begin{equation}
  \sum_{\lambda_3,\lambda_4}
  \big|\mathcal M^{i\,2}\big|^2
  \;\propto\;
  \big|\mathcal M^{(2)}\big|^2_{T}
  + \big|\mathcal M^{(2)}\big|^2_{V}
  + \big|\mathcal M^{(2)}\big|^2_{S},
\end{equation}
for each choice of initial state $i=0,1/2,1$.  In the center-of-mass
frame one can choose $n^\mu=(1,\mathbf 0)$ for convenience, which
identifies $e_L^\mu$ with the standard helicity-0 polarization of the
massive spin-1 building blocks and yields the usual separation into
transverse (helicity $\pm 2$), vector (helicity $\pm 1$) and
scalar (helicity $0$) contributions.

%%%%%%%%%%%%%%%%%%%%%%%%%%%%%%%%%%%%%%%%%%%%%%%%%%%%%%%%%%%%%%%%
\subsection{Gravitational production from the inflaton condensate}
%%%%%%%%%%%%%%%%%%%%%%%%%%%%%%%%%%%%%%%%%%%%%%%%%%%%%%%%%%%%%%%%
We first analyze the production of massive spin-2 particles from the
coherent inflaton condensate.  During reheating, the inflaton field
oscillates around the minimum of its potential and behaves as a gas of matter-like non-relativistic quanta.  Although gravitational interactions are Planck-suppressed, the energy density stored in the condensate is initially very large, so the scattering of inflaton quanta through virtual graviton exchange can provide an efficient source of $X_{\mu\nu}$ quanta. In practice, the production rate is dominated by the earliest stages of reheating, when the oscillation amplitude is maximal and the inflaton energy density is largest.

Our analysis does not rely on a specific inflationary model. It
requires only that the potential admit a well-defined minimum around
which the inflaton oscillates.  For instance, in Starobinsky
inflation~\cite{Starobinsky:1980te} and in $\alpha$-attractor T-models of inflation~\cite{Kallosh:2013hoa} the potential is effectively quadratic near the minimum, so the post-inflationary dynamics are well described by coherent oscillations with an effective mass $m_\phi$.\footnote{For a recent discussion of constraints on T-model inflation and reheating using ACT DR6 and SPT-3G data, see Ref.~\cite{Garcia:2025wmu}.} In the following, we treat this quadratic regime explicitly, and later comment on more general minima.
%%%%%%%%%%%%%%%%%%%%%%%%%%%%%%%%%%%%%%%%%%%%%%%%%%%%%%%%%%%%%%%%
\subsubsection{Quadratic potential minimum}
%%%%%%%%%%%%%%%%%%%%%%%%%%%%%%%%%%%%%%%%%%%%%%%%%%%%%%%%%%%%%%%%
We begin with the simplest case in which the inflaton potential near
its minimum is approximately quadratic,
\begin{equation}
  V(\phi) \;\simeq\; \frac12 m_\phi^2 \phi^2\,.
\end{equation}
In this regime the inflaton condensate can be viewed as a collection
of non-relativistic particles of mass $m_\phi$ at rest in the
condensate frame. The gravitational production process is then
described by the $2\to2$ scattering
\begin{equation}
  \phi(p_1) + \phi(p_2) \;\rightarrow\; X(p_3) + X(p_4)\,,
\end{equation}
with $p_1 = p_2 = (m_\phi,\mathbf{0})$ for the initial inflaton quanta
and $p_3^2 = p_4^2 = m_2^2$ for the final massive spin-2 particles.

The tree-level amplitude mediated by a single graviton exchange,
cf.\ Eq.~\eqref{Eq:scamp_spin2}, can be written schematically as
\begin{equation}
  \mathcal M^{0\,2}
  \;\propto\;
  M^{2}_{\mu\nu}(p_3,p_4)\,
  \Pi^{\mu\nu\rho\sigma}(k)\,
  M^{0}_{\rho\sigma}(p_1,p_2)\,,
\end{equation}
where $k = p_1 + p_2$ is the graviton momentum,
$M^0_{\rho\sigma}$ is the scalar partial tensor in
Eq.~\eqref{Eq:partamp_scalar_spin2}, and $M^{2}_{\mu\nu}$ is the
spin-2 partial tensor given in Eq.~\eqref{Eq:M_spin2_partial}.

Using Eqs.~\eqref{Eq:M_spin2_partial} and
\eqref{eq:spin2_projector}, the polarization-summed squared amplitude
entering the production rate takes the schematic form
\begin{equation}
  \sum_{\lambda_3,\lambda_4}
  \big|\mathcal M^{0\,2}\big|^2
  \;\propto\;
  \Big(
    M^{2}_{\mu\nu}\,
    \Pi^{\mu\nu\rho\sigma}\,
    M^{0}_{\rho\sigma}
  \Big)
  \Big(
    M^{2}_{\alpha\beta}\,
    \Pi^{\alpha\beta\gamma\delta}\,
    M^{0}_{\gamma\delta}
  \Big)^{\!*},
\end{equation}
where the spin-2 polarization sums are encoded in the projector
$\mathcal P_{\mu\nu,\rho\sigma}$ through $M^{2}_{\mu\nu}$.
This expression is sufficient for the unpolarized production rate
from the inflaton condensate, which we will integrate over phase space to obtain the gravitational abundance of $X_{\mu\nu}$ for the quadratic minimum.

In the centre-of-mass frame we take
\begin{equation}
  p_1^\mu = p_2^\mu = (m_{\phi},\mathbf{0}),
  \quad
  p_3^\mu = (E,\mathbf{p}),\quad
  p_4^\mu = (E,-\mathbf{p}),
\end{equation}
with
\begin{equation}
  E = m_{\phi},\qquad
  |\mathbf{p}| = p_3 = \sqrt{m_{\phi}^2 - m_2^2}
  = m_{\phi}\sqrt{1-\tau} \,  ,
\end{equation}
where $ \tau \equiv m_2^2/m_{\phi}^2$ and $0<\tau<1$ is required kinematically.

After including the combinatorial factor $1/2!$ for two identical particles in the initial state and
summing over the five spin-2 polarizations, we obtain for the total (unpolarized) squared matrix element
\begin{equation}
\begin{aligned}
     |\overline{\mathcal{M}}|^2 &\; = \; \frac{1}{288} \frac{m_{\phi}^4}{M_P^4} \tau^{-4} \left(2 + \tau\right)^2 \\
   & \; \times \; \left(64 - 128 \tau + 184 \tau^2 - 120 \tau^3 + 45 \tau^4 \right) \, .
\end{aligned}
\label{eq:M2_total_spin2}
\end{equation}
By construction, the bar indicates that final state combinatorial factors have been taken into account to avoid overcounting.

The associated production rate from the inflaton condensate in a quadratic minimum is obtained by inserting the above result into the standard expression for $2\to2$ annihilations of non-relativistic inflaton quanta,~\cite{Mambrini:2021cwd}~
\begin{equation}
  R^{\phi^2}
  \;\equiv\;
  n_{\phi}^2 \,\langle \sigma v \rangle_{\phi\phi\to XX}
  \; = \; 
  \frac{\rho_{\phi}^2}{m_{\phi}^2}\,
  \frac{|\overline{\mathcal{M}}|^2}
       {16\pi m_{\phi}^2}\,
  \frac{p_3}{m_{\phi}}\,,
  \label{eq:Rphi2_generic}
\end{equation}
with $p_3 = \sqrt{m_{\phi}^2 - m_2^2}$ as above.\footnote{We note that our expression carries a factor of $16\pi$ in the denominator rather than $32\pi$ as in Refs.~\cite{Garcia:2023obw, Kaneta:2023uwi}. This difference is purely conventional: those references include the usual $1/2!$ combinatorial factor associated with identical inflaton quanta in the initial state, whereas we do not, since we do not treat the condensate legs as identical particles in the phase-space bookkeeping. No physical result depends on this choice, it only changes how factors of two are organized.} Substituting
Eq.~\eqref{eq:M2_total_spin2}, we obtain
\begin{equation}
\begin{aligned}
   R^{\phi^2}
   &=
   \frac{2 \times \rho_{\phi}^2}{9216 \pi M_P^4}
   \tau^{-4}\,
   (2 + \tau)^2 \\
   &\times \Big(
     64 - 128 \tau + 184 \tau^2 - 120 \tau^3 + 45 \tau^4
   \Big)\,
   \sqrt{1- \tau} \, .
\end{aligned}
\label{eq:Rphi2_spin2_final}
\end{equation}
Here we included an additional overall factor of $2$  to
explicitly count the two spin-2 particles produced per inflaton
annihilation in the definition of $R^{\phi^2}$.

Using the SVT decomposition of the massive spin-2 projector (see Sec.~\ref{sec:svt}), we can split the result into scalar, vector, and tensor helicity contributions,
\begin{equation}
  |\overline{\mathcal{M}}|^2
  =
  |\overline{\mathcal{M}}|^2_{S}
  +
  |\overline{\mathcal{M}}|^2_{V}
  +
 | \overline{\mathcal{M}}|^2_{T}\,,
\end{equation}
which can be written in the compact form
\begin{align}
  |\overline{\mathcal{M}}|^2_{S}
  &= \frac{1}{288}\,
     \frac{m_{\phi}^4}{M_P^4}\,
     \tau^{-4}\,
     (2 + \tau)^2\,
     \big(8 - 8\tau + 3 \tau^2\big)^2,
  \label{eq:MS_spin2}
  \\[4pt]
  |\overline{\mathcal{M}}|^2_{V}
  &= \frac{1}{16}\,
     \frac{m_{\phi}^4}{M_P^4}\,
     \tau^{-2}\,
     \big(4 - \tau^2\big)^2,
  \label{eq:MV_spin2}
  \\[4pt]
 | \overline{\mathcal{M}}|^2_{T}
  &= \frac{1}{16}\,
     \frac{m_{\phi}^4}{M_P^4}\,
     (2 + \tau)^2.
  \label{eq:MT_spin2}
\end{align}
Here $|\overline{\mathcal{M}}|^2_{S}$ receives contributions from the helicity-0 mode, $|\overline{\mathcal{M}}|^2_{V}$ from the helicity $\pm 1$ modes, and $|\overline{\mathcal{M}}|^2_{T}$ from the
helicity $\pm 2$ modes of $X_{\mu\nu}$, in one-to-one correspondence with the scalar, vector, and tensor projectors introduced previously. A few comments are in order. For $\tau\ll 1$, the three pieces scale as $|\overline{\mathcal{M}}|^2_{T} \sim \mathcal{O}(\tau^{0})$, $|\overline{\mathcal{M}}|^2_{V} \sim \mathcal{O}(\tau^{-2})$, $|\overline{\mathcal{M}}|^2_{S} \sim \mathcal{O}(\tau^{-4})$, so the total amplitude is dominated by the scalar (helicity-$0$) component in the light-mass regime, as anticipated from the SVT analysis.

It is instructive to compare the behavior of Eq.~\eqref{eq:M2_total_spin2}
in the light-mass limit with the familiar scalar, fermion, and vector
cases. For minimally coupled scalars the massless limit of the
gravitational production amplitude is finite, while for fermions there
is no production of strictly massless states due to helicity
conservation, and for massless gauge bosons conformal invariance
forbids gravitational production~\cite{Mambrini:2021zpp,Clery:2021bwz,
Ema:2015dka,Ema:2016hlw,Ema:2018ucl,Ema:2019yrd,Garcia:2023obw}.
By contrast, the massive spin-2 amplitude in
Eq.~\eqref{eq:M2_total_spin2} grows as $\tau^{-4}$, reflecting the
well-known pathologies associated with higher-spin fields coupled to
gravity~\cite{Kolb:2021xfn,Kolb:2023dzp,Kaneta:2023uwi}.

In particular, the tensor contribution
$|\overline{\mathcal{M}}|^2_{T}$ is the direct analogue of the
gravitational production of two minimally coupled scalar degrees of
freedom, but with two transverse tensor polarizations instead of a
single real scalar. Since there are two such polarizations, the tensor
contribution is enhanced by a factor of two compared to the production
of a single real scalar through the same gravitational channel
(see, e.g., Refs.~\cite{Mambrini:2021zpp,Clery:2021bwz} for the scalar case).
The strong $\tau^{-2}$ and $\tau^{-4}$ enhancements are entirely due
to the helicity-$1$ and helicity-$0$ sectors, respectively. We emphasize that the formal $\tau \to 0$ limit lies outside the regime of validity of the Fierz-Pauli effective theory and should be interpreted as signaling the onset of strong coupling in the longitudinal sector, rather than as a physical divergence in the
production of massless gravitons.

We note that in processes such as $\phi\phi\to X_{\mu\nu}X_{\mu\nu}$ with a \emph{massive} spin-2 final state, the apparent inverse-mass enhancement originates from the longitudinal/helicity-0 sector: for $E\gg m_2$, the helicity-0 polarization tensor contains pieces scaling schematically as $\epsilon^{(0)}_{\mu\nu}\sim p_\mu p_\nu/m_2^2$, so amplitudes can grow rapidly with energy and yield parametric behavior as strong as $|{\cal M}|^2\propto (E/m_2)^8$ in a naive Fierz-Pauli effective theory. This rapid growth is the same physics behind the low strong-coupling cutoff of massive gravity, e.g.\ $\Lambda_5\sim (M_{\rm Pl} m_2^4)^{1/5}$, indicating the breakdown of the minimal EFT once $E$ approaches the cutoff~\cite{Arkani-Hamed:2002bjr, Aubert:2003je, Hinterbichler:2011tt}

%%%%%%%%%%%%%%%%%%%%%%%%%%%%%%%%%%%%%%%%%%%%%%%%%%%%%%%%%%%%%
\subsubsection{General potentials}
\label{subsec:general_potentials_spin2}
%%%%%%%%%%%%%%%%%%%%%%%%%%%%%%%%%%%%%%%%%%%%%%%%%%%%%%%%%%%%%
As discussed above, the gravitational production of massive spin-2 particles from the inflaton condensate depends sensitively on the shape of the potential near its minimum.  The derivation of the Boltzmann equation for a decaying inflaton condensate follows the same steps as in Appendix~\ref{app:boltz}, and we now extend that treatment to the case of a Fierz-Pauli field $X_{\mu\nu}$.

We consider general single-field potentials of the form
\begin{equation}
    V(\phi) \; = \; \lambda M_P^4 \left(\frac{\phi}{M_P} \right)^k \, , \qquad \phi \ll M_P \, ,
    \label{eq:genpot}
\end{equation}
where $k \geq 2$, and for which the inflaton oscillates periodically about a minimum at $\phi_0$.  Using Eq.~\eqref{Eq:oscillation}, the
time-dependence of the background can be parametrized as
$V(\phi) = V(\phi_0)\,\mathcal{P}(t)^k$, where $\mathcal{P}(t)$ is a
dimensionless periodic function.  Fourier expanding $\mathcal{P}(t)^k$
as in Eq.~\eqref{PFexpand}, the potential can be written as a sum over
harmonics of the fundamental oscillation frequency,~\cite{GKMO2, Shtanov:1994ce, Ichikawa:2008ne, Kainulainen:2016vzv}
\begin{equation}
  V(\phi)
  = V(\phi_0)\sum_{n=-\infty}^{\infty} \mathcal{P}_{k,n}\,e^{-i n \omega t}
  = \langle \rho_\phi \rangle
    \sum_{n=-\infty}^{\infty} \mathcal{P}_{k,n}\,e^{-i n \omega t}\,,
\end{equation}
where $\langle \rho_\phi \rangle$ denotes the inflaton energy density averaged over many oscillations and $\omega$ is the oscillation frequency, given by~\cite{GKMO2}
\begin{equation}
  \label{eq:angfrequency_spin2}
  \omega
  \;=\;
  m_\phi
  \sqrt{\frac{\pi k}{2(k-1)}}
  \frac{\Gamma\!\left(\tfrac{1}{2}+\tfrac{1}{k}\right)}
       {\Gamma\!\left(\tfrac{1}{k}\right)}\,,
\end{equation}
with
$m_\phi^2 \equiv \partial^2 V / \partial \phi^2 |_{\phi_0}$.

Each Fourier mode with frequency $E_n \equiv n\omega$ acts as a coherent source for annihilations of two inflaton quanta with energy $E_n/2$ into a pair of massive spin-2 particles, with $E_n = n\omega \ge 2 m_2$. Following the procedure outlined in Appendix~\ref{app:boltz}, and using the spin-2 matrix element derived in the previous section, the gravitational production rate from the inflaton condensate can be written as
\begin{equation}
    \label{eq:rategenk_spin2}
    R^{\phi^k}
    \;\equiv\;
    n_\phi^2 \langle \sigma v \rangle_{\phi\phi\to XX}
    \;=\;
    \frac{2 \times \rho_{\phi}^2}{576 \pi M_P^4}\,\Sigma_{2}^k\,,
\end{equation}
with
\begin{equation}
\begin{aligned}
    \Sigma_2^k & \; = \; \sum_{n = 1}^{\infty} |\mathcal{P}_{k, n}|^2 \frac{E_n^8}{m_2^8}\left(1 + 2\frac{m_2^2}{E_n^2} \right)^2 \sqrt{1- \frac{4m_{2}^2}{E_n^2}}\\
    & \; \times \; \left[1 - 8\frac{m_2^2}{E_n^2} + 46 \frac{m_2^4}{E_n^4} - 120 \frac{m_2^6}{E_n^6} + 180 \frac{m_2^8}{E_n^8}\right]  \, ,
 \end{aligned}   
\end{equation}
where the factor of $2$ explicitly counts the two spin-2 particles produced per annihilation, and $\Sigma_{2}^k$ encodes the dependence on the inflaton potential and on the spin-2 mass. In the quadratic limit $k=2$, the inflaton oscillates as $\mathcal{P}(t)^2 = \cos^2(m_\phi t)
  = \frac12 + \frac14(e^{-2 i m_\phi t} + e^{2 i m_\phi t})$,
so only the $n=\pm2$ harmonics contribute, and
\begin{equation}
  \sum_{n}|\mathcal{P}_{2,n}|^2 = |\mathcal{P}_{2,2}|^2 = \frac{1}{16} \, .
  \label{eq:pdef}
\end{equation}
Using $E_2 = 2 m_\phi$ one finds that Eqs.~\eqref{eq:rategenk_spin2}-\eqref{eq:pdef} reproduce the quadratic result \eqref{eq:Rphi2_spin2_final}.

We can decompose the total rate in Eq.~\eqref{eq:rategenk_spin2} into
its scalar, vector, and tensor helicity contributions as
\begin{equation}
    \label{eq:rategenk_spin2general}
    R^{\phi^k}
    \;\equiv\;
    n_\phi^2 \langle \sigma v \rangle_{\phi\phi\to XX}
    \;=\;
    \frac{2 \times \,\rho_{\phi}^2}{576 \pi M_P^4}\,
    \left( \Sigma_{2,S}^k + \Sigma_{2,V}^k + \Sigma_{2,T}^k \right)\,,
\end{equation}
where the scalar (helicity-0) contribution is
\begin{equation}
\begin{aligned}
    \label{eq:ratescalar}
     \Sigma_{2,S}^k
     \; = \;
     \sum_{n=1}^{\infty}
     &|\mathcal{P}_{k, n}|^2
     \,\frac{E_n^8}{m_2^8}
     \left(1 + 2 \frac{m_2^2}{E_n^2}\right)^2 \\
     &\times
     \left(1 - 4 \frac{m_2^2}{E_n^2}
              + 6 \frac{m_2^4}{E_n^4} \right)^2
     \left[1 - \frac{4m_2^2}{E_n^2} \right]^{1/2} \, ,
\end{aligned}
\end{equation}
the vector (helicity $\pm 1$) contribution is
\begin{equation}
\begin{aligned}
    \label{eq:ratevector}
     \Sigma_{2,V}^k
     \; = \;
     \sum_{n=1}^{\infty}
     &|\mathcal{P}_{k, n}|^2
     \times 18\,
     \frac{E_n^4}{m_2^4}
     \left(1 - 4 \frac{m_2^4}{E_n^4}\right)^2
     \left[1 - \frac{4m_2^2}{E_n^2} \right]^{1/2}  \, ,
\end{aligned}
\end{equation}
and the tensor (helicity $\pm 2$) contribution is
\begin{equation}
\begin{aligned}
     \label{eq:ratetensor}
     \Sigma_{2,T}^k
     \; = \;
     \sum_{n=1}^{\infty}
     &|\mathcal{P}_{k, n}|^2
     \times 72\,
     \left(1 + 2\frac{m_2^2}{E_n^2}\right)^2
     \left[1 - \frac{4m_2^2}{E_n^2} \right]^{1/2}  \, .
\end{aligned}
\end{equation}
Only the tensor contribution admits a smooth massless limit
$m_2 \rightarrow 0$, as expected from the discussion in
the previous section.

\begin{figure}[h!]
    \centering
    \includegraphics[width=1\linewidth]{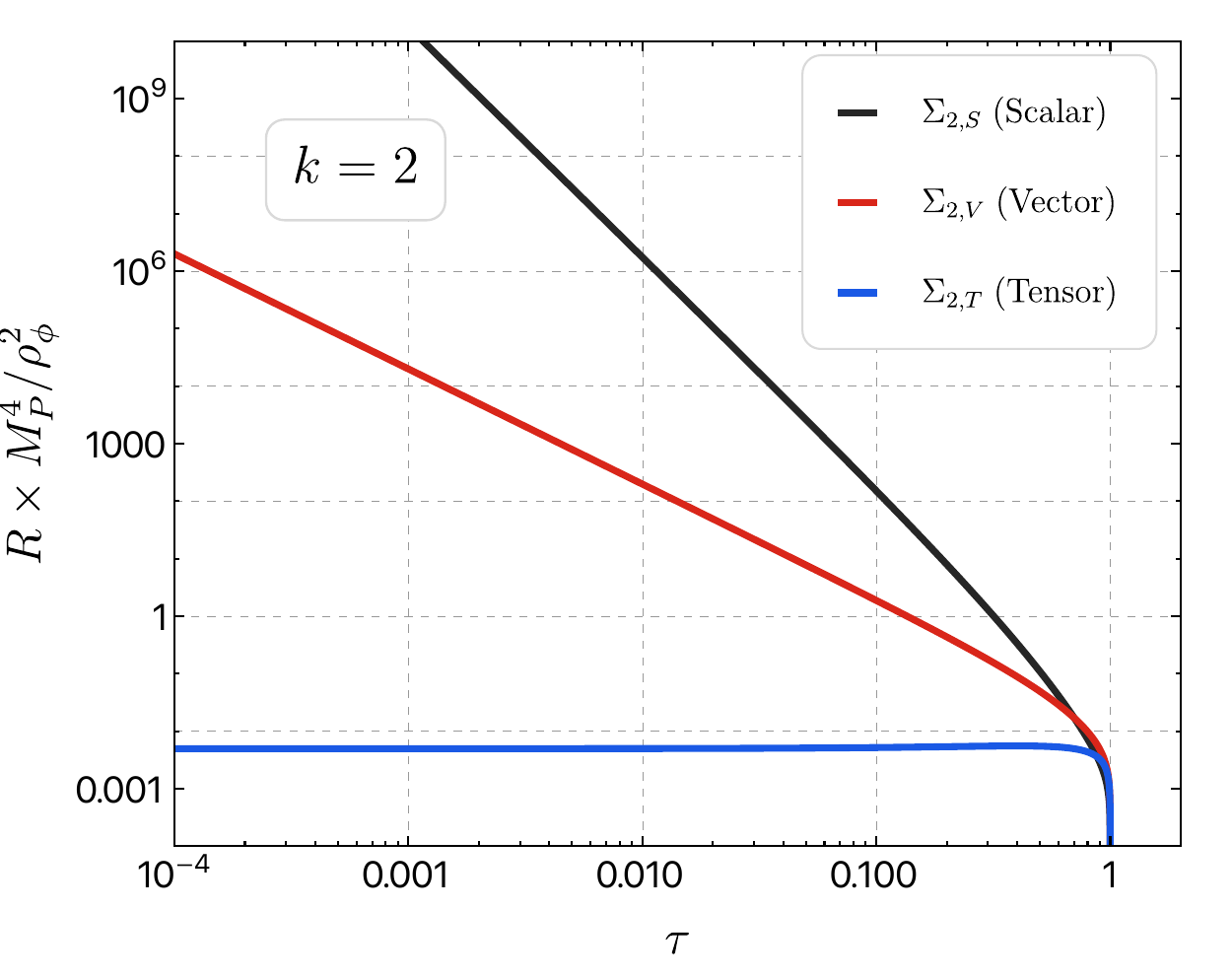}
    \caption{Scalar (black), vector (red), and tensor (blue) contributions to the gravitational production rates of a massive spin-2 particle for $k = 2$ in units of $R \times M_P^4/\rho_{\phi}^2$ as a function of $\tau = m_2^2/m_{\phi}^2$.}
    \label{fig:svtrates}
\end{figure}

The behavior of these contributions is illustrated in
Fig.~\ref{fig:svtrates}, which shows the scalar
(Eq.~\eqref{eq:ratescalar}), vector
(Eq.~\eqref{eq:ratevector}), and tensor
(Eq.~\eqref{eq:ratetensor}) rates as functions of
$\tau = m_2^2/m_{\phi}^2$ for a quadratic inflaton potential ($k = 2$). As anticipated from the helicity analysis, the scalar (helicity-0) piece, which scales as $\propto \tau^{-4}$ in the light-mass regime, always dominates over the vector contribution, whose leading behavior is $\propto \tau^{-2}$, while the tensor part
remains $\mathcal{O}(\tau^0)$ and is therefore subdominant for $\tau \ll 1$. Already for $\tau = 1/2$ we find that roughly two thirds of the total production rate is carried by the scalar helicity-0 component.

This pattern mirrors what is found for massive spin-1 and spin-3/2 fields, where the longitudinal mode dominates the gravitational production rate in the light-mass regime
\cite{Garcia:2023obw,Kolb:2020fwh,Kaneta:2023uwi,Kolb:2021xfn}. Similar numerical results for the spin-2 case were obtained in Ref.~\cite{Kolb:2023dzp}, where the scalar component was likewise found to dominate over the vector and tensor helicities.

    Finally, we emphasize that the overall normalization of the rate is controlled by the Fourier coefficients $\mathcal{P}_{k,n}$ of the inflaton oscillations. For large $k$ these coefficients become increasingly similar across different powers of the potential, so we do not expect large qualitative differences in the relative importance of the scalar, vector, and tensor contributions when varying $k$.

%%%%%%%%%%%%%%%%%%%%%%%%%%%%%%%%%%%%%%%%%%%%%%%%%%%%%%%%%%%%%%%%
\section{Relic Abundance Computation}
\label{sec:abundance}
%%%%%%%%%%%%%%%%%%%%%%%%%%%%%%%%%%%%%%%%%%%%%%%%%%%%%%%%%%%%%%%%
To determine the relic abundance of the massive spin-2 field, we begin
with the Boltzmann equation for its number density $n$,
\begin{equation}
  \frac{dn}{dt} + 3Hn = R^{\phi^k} \,,
  \label{Eq:boltzmann1}
\end{equation}
where $R^{\phi^k}$ denotes the gravitational production rate sourced
by the inflaton condensate for a potential of the form $V(\phi)\propto \phi^k$. Introducing the comoving yield $Y \equiv a^3 n$, Eq.~\eqref{Eq:boltzmann1} can be rewritten as a
first-order equation in the scale factor,
\begin{equation}
  \frac{dY}{da}
  \;=\; \frac{a^2 R^{\phi^k}}{H(a)} \, .
  \label{Eq:boltzmann2}
\end{equation}

The same interactions that produce the spin-2 field also transfer
energy from the inflaton condensate to the radiation bath. At the
background level this is captured by the coupled evolution equations
\begin{align}
  \dot{\rho}_{\phi} + 3H(1+w_{\phi})\rho_{\phi}
  &\;\simeq\;
  -\Gamma_{\phi}(1+w_{\phi})\rho_{\phi} \, ,
  \label{rhoeom1} \\
  \dot{\rho}_{R} + 4H \rho_{R}
  &\;\simeq\;
  (1+w_{\phi})\Gamma_{\phi}(t)\rho_{\phi}\, ,
  \label{aeom}
\end{align}
where $H=\dot a/a$ is the Hubble parameter and $w_{\phi} = P_\phi/\rho_\phi$ is the inflaton equation of state
parameter. For an inflaton oscillating in a monomial potential
$V(\phi)\propto \phi^k$, one has~\cite{GKMO2, Turner:1983he}
\begin{equation}
  \label{eq:wk}
  w_{\phi} \; = \; \frac{k-2}{k+2} \,.
\end{equation}
Using $\tfrac{d}{dt} = a H \tfrac{d}{da}$, Eq.~\eqref{rhoeom1} can be integrated (neglecting the subdominant decay term in the early stages of reheating) to obtain
\begin{equation}
  \rho_\phi(a)
  = \rho_{\rm end}
    \left(\frac{a}{a_{\rm end}} \right)^{-\frac{6k}{k+2}} \, .
  \label{Eq:rhophi}
\end{equation}
Here $\rho_{\rm end} \equiv \rho_\phi(a_{\rm end})$ is the inflaton energy density at the end of inflation, where $a_{\rm end}$ is defined by $\phi(a_{\rm end}) = \phi_{\rm end}$ and $\phi_{\rm end}$ is the field value at the end of inflation, i.e., when $\ddot{a} = 0$, implying $\rho_{\rm end} = \tfrac32 V(\phi_{\rm end})$.

Throughout reheating, we assume that the inflaton energy density dominates the total energy density, so that
\begin{equation}
  H(a)
  \; = \; \frac{\rho_\phi^{1/2}(a)}{\sqrt{3}\,M_P} \, .
\end{equation}
Using Eq.~\eqref{Eq:rhophi}, the Boltzmann equation
\eqref{Eq:boltzmann2} can be written as
\begin{equation}
  \frac{dY}{da}
  = \frac{\sqrt{3}\,M_P}{\sqrt{\rho_{\rm RH}}}
    a^2
    \left(\frac{a}{a_{\rm RH}}\right)^{\frac{3k}{k+2}}
    R^{\phi^k}(a) \, ,
  \label{Eq:boltzmann4}
\end{equation}
where $\rho_{\rm RH}$ and $a_{\rm RH}$ denote the total energy density and scale factor at (the end of) reheating.

We first focus on the simplest case $k = 2$, for which the inflaton oscillates in an approximately quadratic minimum. In this case, $\rho_\phi \propto a^{-3}$, while the radiation energy density scales as $\rho_R \propto T^4 \propto a^{-3/2}$ during reheating, and the inflaton mass is $m_\phi^2 = 2 \lambda M_P^2$.  Using the spin-2 gravitational production rate $R^{\phi^2}(a)$ given by
Eq.~\eqref{eq:rategenk_spin2}, Eq.~\eqref{Eq:boltzmann4} can be integrated analytically. Evaluating the result at reheating and expressing it in terms of the reheating temperature $T_{\rm RH}$,
\begin{equation}
\begin{aligned}
  n(a_{\rm RH})
  &\; =\;
  \frac{1}{2304 \sqrt{3} \pi  M_P}
  \left(\frac{\rho_{\rm end}}{M_P^4} \right)^\frac12
  \frac{g_{\rm RH} \pi^2}{30} T_{\rm RH}^4
  \,\tau^{-4} \left(2 + \tau\right)^2
  \\
  & \quad \times
  \left(64 - 128 \tau + 184 \tau^2 - 120 \tau^3 + 45 \tau^4 \right)
  \sqrt{1 - \tau} \, ,
  \label{n2tot}
\end{aligned}
\end{equation}
where $\tau \equiv m_2^2/m_\phi^2$, we have assumed
$a_{\rm RH} \gg a_{\rm end}$ and denoted by $g_{\rm RH}$ the number of relativistic degrees of freedom at reheating.

The present-day relic abundance can be expressed as~\cite{Mambrini:2021cwd}
\begin{equation}
  \label{eq:relicabund1}
  \Omega h^2
  \;\simeq\;
  1.6\times 10^8
  \frac{g_0}{g_{\rm RH}}
  \frac{n(T_{\rm RH})}{T_{\rm RH}^3}
  \frac{m_{2}}{1~{\rm GeV}} \, ,
\end{equation}
where $g_0$ is the effective number of entropy degrees of freedom
today.  Using Eq.~\eqref{n2tot}, and normalizing to representative
inflationary parameters, we find
\begin{equation}
\begin{aligned}
  \Omega h^2
  & \simeq  0.12 \times
  \left( \frac{T_{\rm RH}}{6 \, {\rm GeV}} \right)
  \left( \frac{\rho_{\rm end}}{(5.2 \times 10^{15} {\rm GeV})^4} \right)^\frac12
  \\
  & \quad \times
  \left( \frac{m_{\phi}}{1.7 \times 10^{13} \, \rm GeV} \right)^8
  \left( \frac{8 \times 10^{12} \, \rm GeV}{m_2} \right)^7 \, ,
  \label{oh2tot}
\end{aligned}
\end{equation}
where we have used $g_0=43/11$ and $g_{\rm RH}=427/4$, and assumed $m_{2} \ll m_\phi$. The reference values for $m_\phi$ and $\rho_{\rm end}$ correspond to a typical $\alpha$-attractor model with $k=2$, though the precise
numerical normalization depends only mildly on $T_{\rm RH}$ (see Refs.~\cite{GKMO1,Clery:2021bwz} for a detailed discussion).

Comparing the gravitational production of massive spin-2 particles in Eq.~\eqref{oh2tot} with that of scalars
\cite{Mambrini:2021zpp,Clery:2021bwz}, fermions
\cite{Mambrini:2021zpp,Clery:2021bwz}, vectors~\cite{Garcia:2023obw}, and spin-3/2 raritrons~\cite{Kaneta:2023uwi}, we see that spin-2
production is parametrically far more efficient because the total matrix element squared scales as $\tau^{-4}$ in the light-mass limit. Raritron production is already strongly enhanced by a factor $\sim m_\phi^2/m_{3/2}^2$, but the spin-2 amplitude exhibits an even stronger divergence in the $\tau\ll 1$ regime. Consequently, for
fixed inflaton parameters, massive spin-2 dark matter typically overcloses the Universe unless either the reheating temperature $T_{\rm RH}$ is quite low or the spin-2 mass lies close to the kinematic threshold, $2m_2 \simeq m_{\phi}$.

We can now decompose the total abundance into scalar, vector, and tensor helicity contributions using the rates in Eqs.~\eqref{eq:ratescalar}-\eqref{eq:ratetensor}.  For the scalar component with $k=2$, the number density at reheating is
\begin{equation}
\begin{aligned}
    n_S(a_{\rm RH})
    &\; = \;
    \frac{1}{2304 \sqrt{3} \pi  M_P}
    \left(\frac{\rho_{\rm end}}{M_P^4} \right)^\frac12
    \frac{g_{\rm RH} \pi^2}{30} T_{\rm RH}^4
    \,\tau^{-4} 
    \\
    &\times
    \left(8 - 8\tau + 3 \tau^2\right)^2
    \left(2 + \tau\right)^2 \sqrt{1 - \tau} \, ,
\end{aligned}
\end{equation}
leading to
\begin{equation}
\begin{aligned}
  \Omega_S h^2
  & \simeq 0.12 \times
  \left( \frac{T_{\rm RH}}{6 \, {\rm GeV}} \right)
  \left( \frac{\rho_{\rm end}}{(5.2 \times 10^{15} {\rm GeV})^4} \right)^\frac12
  \\
  &\quad \times
  \left( \frac{m_{\phi}}{1.7 \times 10^{13} \, \rm GeV} \right)^8
  \left( \frac{8 \times 10^{12} \, \rm GeV}{m_2} \right)^7 \, ,
  \label{oh2totscalar}
\end{aligned}
\end{equation}
which coincides with the total abundance in Eq.~\eqref{oh2tot},
confirming that the relic density is dominated by the helicity-0 mode.

For the vector-like helicity sector, we obtain
\begin{equation}
\begin{aligned}
    n_V(a_{\rm RH})
    &\; = \;
    \frac{1}{128 \sqrt{3} \pi  M_P}
    \left(\frac{\rho_{\rm end}}{M_P^4} \right)^\frac12
    \frac{g_{\rm RH} \pi^2}{30} T_{\rm RH}^4
    \\
    &\quad \times
    \tau^{-2} \left(4- \tau^2\right)^2
    \sqrt{1 - \tau} \, ,
\end{aligned}
\end{equation}
and
\begin{equation}
\begin{aligned}
  \Omega_V h^2
  & \simeq  9.1 \times 10^{-3} \times
  \left( \frac{T_{\rm RH}}{6 \, {\rm GeV}} \right)
  \left( \frac{\rho_{\rm end}}{(5.2 \times 10^{15} {\rm GeV})^4} \right)^\frac12
  \\
  &\quad \times
  \left( \frac{m_{\phi}}{1.7 \times 10^{13} \, \rm GeV} \right)^4
  \left( \frac{8 \times 10^{12} \, \rm GeV}{m_2} \right)^3 \, ,
  \label{oh2totvector}
\end{aligned}
\end{equation}
while the tensor component gives
\begin{equation}
\begin{aligned}
    n_T(a_{\rm RH})
    &\; = \;
    \frac{1}{128 \sqrt{3} \pi  M_P}
    \left(\frac{\rho_{\rm end}}{M_P^4} \right)^\frac12
    \frac{g_{\rm RH} \pi^2}{30} T_{\rm RH}^4
    \\
    &\quad \times
    \left(2 + \tau \right)^2
    \sqrt{1 - \tau} \, ,
\end{aligned}
\end{equation}
and
\begin{equation}
\begin{aligned}
  \Omega_T h^2
  & \simeq  1.1 \times 10^{-4} \times
  \left( \frac{T_{\rm RH}}{6 \, {\rm GeV}} \right)
  \left( \frac{\rho_{\rm end}}{(5.2 \times 10^{15} {\rm GeV})^4} \right)^\frac12
  \\
  &\quad \times
  \left( \frac{8 \times 10^{12} \, \rm GeV}{m_2} \right) \, .
  \label{oh2tottensor}
\end{aligned}
\end{equation}
As expected from the $\tau$ scaling of the helicity amplitudes, both the vector and tensor contributions are negligible compared to the scalar helicity-0 component in the parameter space of interest.

Finally, we generalize these results to the case $k \neq 2$. For a general monomial potential $V(\phi)\propto \phi^k$, the number density at reheating can be written as
\begin{equation}
  n(a_{\rm RH})
  \; = \;
  \frac{(k+2)}{(k-1)}
  \frac{\rho_{\rm RH}^{3/2}}{576\sqrt{3} \pi M_P^{3}}
  \left(\frac{\rho_{\rm end}}{\rho_{\rm RH}}\right)^{1-\frac{1}{k}}
  \Sigma_{2}^k \, ,
\end{equation}
where $\Sigma_{2}^k$ is defined in Eq.~\eqref{eq:rategenk_spin2}. The corresponding dark matter abundance is
\begin{align}
  \Omega h^2
  &\simeq
  2.2 \times 10^{17}
  \frac{(k+2)}{(k-1)}
  \left(\frac{\rho_{\rm end}}{\rho_{\rm RH}} \right)^{1-\frac{1}{k}} \\
 &  \times  \frac{\rho_{\rm RH}^{3/4}}{M_P^3}
  \left(\frac{m_{2}}{8 \times 10^{12}~{\rm GeV}}  \right)
  \Sigma_{2}^k  \, .
  \label{oh2totgenk2}
\end{align}
For $k=2$ this expression reduces to Eq.~\eqref{oh2tot}.
%%%%%%%%%%%%%%%%%%%%%%%%%%%%%%%%%%%%%%%%%%%%%%%%%%%%%%%%%%%%%%%%
\section{Thermal Production of Massive Spin-2 Dark Matter}
\label{sec:spin2dmthermal}
%%%%%%%%%%%%%%%%%%%%%%%%%%%%%%%%%%%%%%%%%%%%%%%%%%%%%%%%%%%%%%%%

A second production mechanism for massive spin-2 dark matter operates via scatterings in the Standard Model thermal plasma, $\mathrm{SM} + \mathrm{SM} \rightarrow X_{\mu\nu} + X_{\mu\nu}$, also represented in Fig.~\ref{fig:spin2portal}. In this case the partial tensors $M^{i}_{\mu\nu}$ entering the graviton-exchange amplitude \eqref{Eq:M_spin2_partial} are those in Eqs.~\eqref{Eq:partamp_scalar_spin2} -\eqref{Eq:partamp_vector_spin2}, with $i = 0, 1/2, 1$ corresponding to scalar, fermion and vector initial states. The relevant SM particles in the initial state are the Higgs scalars, gauge bosons and fermions. At the beginning of reheating the typical centre-of-mass energies are of order $m_\phi$, much larger than the electroweak scale, so it is an excellent approximation to treat all SM species as massless. A detailed derivation of the amplitudes and the resulting thermal production rate is given in Appendix~\ref{ap:thermal_spin2}.

For scalar initial states we employ Eq.~\eqref{fullM2} (neglecting SM masses compared to the reheating temperature). The corresponding expressions for massless fermions and gauge bosons are given in Eqs.~\eqref{M1232} and~\eqref{M132}, respectively. The total thermal production rate is obtained by integrating the spin- and
species-summed squared amplitude over the phase space of the incoming states \cite{Dudas:2017rpa, Mambrini:2021cwd, Benakli:2017whb},
\begin{equation}
\begin{aligned}
  R(T)
  \; =\;
  \frac{2}{1024 \pi^6} & \times
  \int f_1 f_2\,
       E_1\,\mathrm{d}E_1\,
       E_2\,\mathrm{d}E_2\,
       \mathrm{d}\cos\theta_{12}\, \\
       & \times \int
       \bigl|\overline{\mathcal M}\bigr|^2\,
       \mathrm{d}\Omega_{13} \, ,
  \label{thrate}
 \end{aligned} 
\end{equation}
where we assume $s \gg 4 m_2^2$ with $s = (p_1 + p_2)^2$.  The
overall factor of 2 accounts for the two spin-2 particles produced per
scattering.  Here $E_i$ denote the energies of the initial and final
state particles, $\theta_{13}$ and $\theta_{12}$ are the angles
between $\mathbf p_1$ and $\mathbf p_3$ in the centre-of-mass
frame and between $\mathbf p_1$ and $\mathbf p_2$ in the
laboratory frame, respectively, and $\mathrm{d}\Omega_{13} = 2\pi\,\mathrm{d}\cos\theta_{13}$. The incoming particles follow thermal distributions
\begin{equation}
  f_i \;=\; \frac{1}{e^{E_i/T} \pm 1} \, ,
\end{equation}
and the spin- and species-averaged squared amplitude can be decomposed as
\begin{equation}
  \label{Eq:ampscat}
  \bigl|\overline{\mathcal{M}}\bigr|^{2}
  \;=\;
  4\,\bigl|\overline{\mathcal{M}}^{0}\bigr|^{2}
  + 45\,\bigl|\overline{\mathcal{M}}^{1/2}\bigr|^{2}
  + 12\,\bigl|\overline{\mathcal{M}}^{1}\bigr|^{2} \, ,
\end{equation}
corresponding to $N_b = 4$ real Higgs degrees of freedom,
$N_f = 45$ fermionic degrees of freedom, and $N_V = 12$ gauge
bosons in the SM.  The explicit matrix elements and their derivation
are summarized in Appendix~\ref{ap:thermal_spin2}.  Carrying out the
thermal and angular integrations one arrives at the compact
expression
\begin{equation}
\begin{aligned}
  \label{eq:rate2_main}
  R^{T}_{2}(T)
  &\;=\;
  \beta_{1}\,\frac{T^{16}}{m_{2}^{8} M_{P}^{4}}
  +\beta_{2}\,\frac{T^{14}}{m_{2}^{6} M_{P}^{4}}
  +\beta_{3}\,\frac{T^{12}}{m_{2}^{4} M_{P}^{4}}
  \\
  &\quad
  + \beta_{4}\,\frac{T^{10}}{m_{2}^{2} M_{P}^{4}}
  +\beta_{5}\,\frac{T^{8}}{M_{P}^{4}}
  +\beta_{6}\,\frac{m_{2}^{2}T^{6}}{M_{P}^{4}}
  \\
  &\quad
  + \beta_{7}\,\frac{m_{2}^{4}T^{4}}{M_{P}^{4}} \, ,
\end{aligned}
\end{equation}
where the numerical coefficients $\beta_i$ are given explicitly in
Appendix~\ref{ap:thermal_spin2}.

The rate $R^{T}_{2}(T)$ sources spin-2 production from the thermal bath. For concreteness, we focus on the quadratic case $k = 2$ and compare this contribution with production from the inflaton condensate. Following the same procedure as in the previous section, we insert the thermal rate
\eqref{eq:rate2_main} into the Boltzmann equation
\eqref{Eq:boltzmann4}, replacing $R^{\phi^k}$ by $R^{T}_{2}(T)$. The temperature as a function of the scale factor is obtained from the radiation energy-density evolution,
\begin{equation}
  \frac{\mathrm{d} \rho_R}{\mathrm{d}a}
  + 4\frac{\rho_R}{a}
  \;=\;
  \frac{\Gamma_\phi \rho_\phi}{H a}\,,
\end{equation}
which we solve in the inflaton-dominated era between
$a_{\rm end}$ and $a_{\rm RH}$. Keeping only the leading term in Eq.~\eqref{eq:rate2_main}, we find that the thermally produced spin-2 number density at reheating is
\begin{equation}
  n^{T}(T_{\rm RH})
  \; \simeq \;
  \frac{2 \beta_{1}}{\sqrt{3}\,\alpha^{4}}\,
  \frac{\rho_{\rm RH}^{7/2}}{m_{2}^{8} M_P^{3}} \, ,
  \label{Eq:nthermal}
\end{equation}
where $ \alpha \equiv g_{\rm RH} \pi^2/30$ and we have assumed $4 m_{2}^{2} \ll s \sim T^{2}$, corresponding approximately to $m_{2} \lesssim T_{\rm RH}$. In this regime the first term in Eq.~\eqref{eq:rate2_main} dominates, since $\beta_1 \simeq 6.9\times 10^{3}$ is numerically much larger than $\beta_{i=2,\dots,7}$.  The integration is performed from $a_{\rm end}$ to $a_{\rm RH}$. When $T_{\rm RH}<m_2$ the upper limit is instead the scale factor $a_{2}$ at which $T = m_2$.

Using Eq.~\eqref{eq:relicabund1} for the relic abundance, we obtain
\begin{equation}
\begin{aligned}
  \Omega^{T} h^2
  &= 5.9 \times 10^{6}\,
     \frac{n^{T}(T_{\rm RH})}{T_{\rm RH}^3}
     \frac{m_{2}}{1~{\rm GeV}}
  \\
  &\simeq
  2.7 \times 10^{-4}
  \left( \frac{T_{\rm RH}}{10^{12}~ {\rm GeV}} \right)^{11}
  \left(\frac{8 \times 10^{12} \, \rm{GeV}}{m_2} \right)^7 \, ,
  \label{Oh2therm}
\end{aligned}
\end{equation}
where in the second line we have kept only the dominant contribution proportional to $\beta_1$. Throughout the parameter space of interest the relic density generated by thermal scatterings is negligible compared to the contribution from inflaton condensate annihilations. This conclusion remains unchanged for general monomial minima with $k>2$, since the condensate channel continues to dominate over the thermal one for all cases considered in this work.

\begin{figure}[h!]
    \centering
    \includegraphics[width=0.9\linewidth]{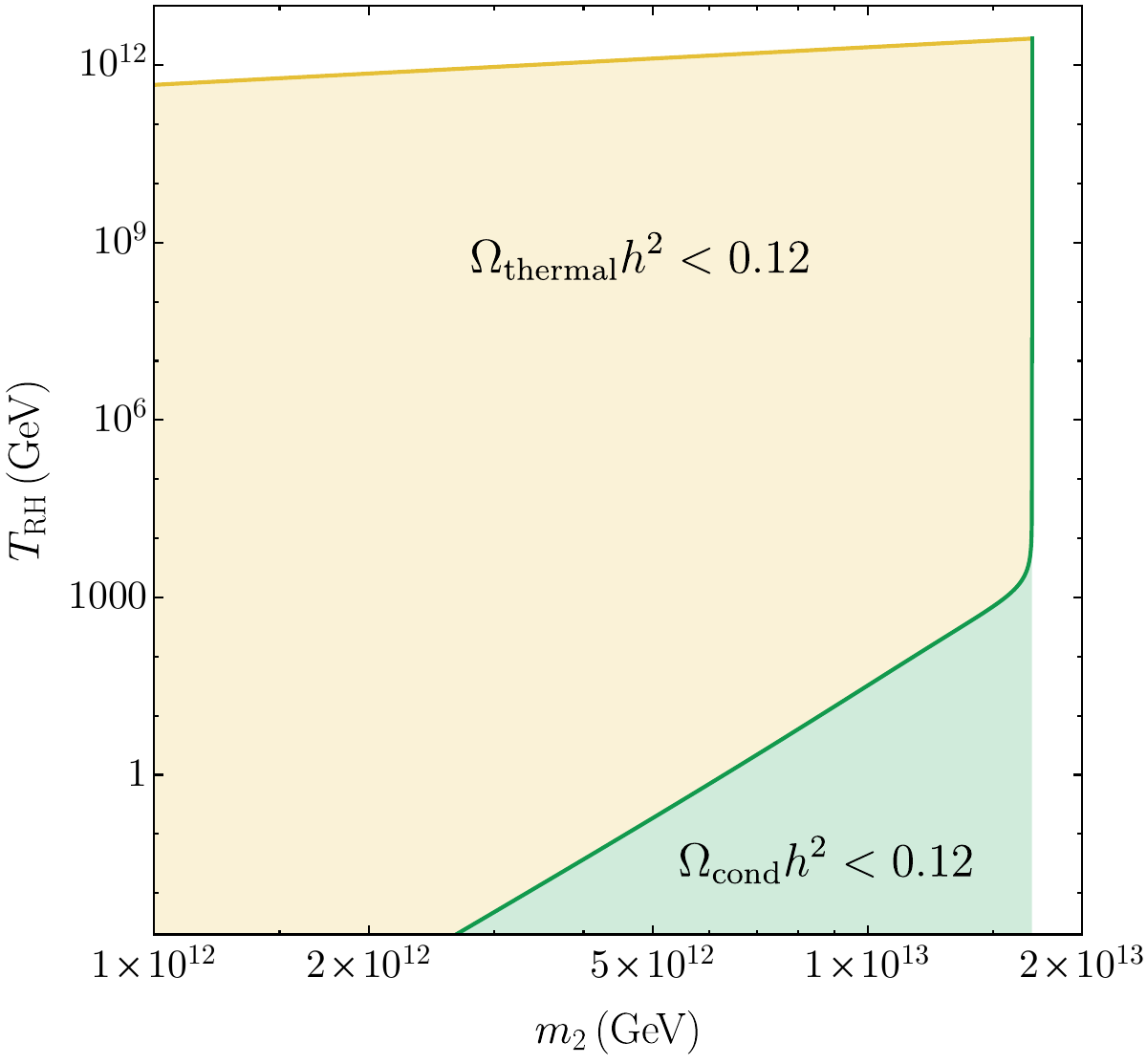}
    \caption{Contours of $\Omega_{\rm cond} h^2 = 0.12$ (green) and
    $\Omega_{\rm thermal} h^2 = 0.12$ (yellow) in the $(m_2, T_{\rm RH})$ plane.}
    \label{fig:dmconst}
\end{figure}

We compare the condensate contribution to the dark matter relic abundance, given by Eq.~(\ref{oh2tot}), with the thermally produced contribution, given by Eq.~(\ref{Oh2therm}), in Fig.~\ref{fig:dmconst}. To remain consistent with Big Bang Nucleosynthesis, we impose a lower bound on the reheating temperature of $T_{\rm RH} \gtrsim T_{\rm BBN} \simeq 2~\mathrm{MeV}$. In our scenario, this constraint implies that the spin-2 dark matter particle cannot be lighter than $m_2 \gtrsim 2.7 \times 10^{12}~\mathrm{GeV}$. This channel is unavoidable: because it is purely gravitational, the inflaton condensate always annihilates into pairs of massive spin-2 particles at some level.

%%%%%%%%%%%%%%%%%%%%%%%%%%%%%%%%%%%%%%%%%%%%%%%%%%%%%%%%%%%%%%%%
\section{Summary and Conclusions}
\label{sec:summary}
%%%%%%%%%%%%%%%%%%%%%%%%%%%%%%%%%%%%%%%%%%%%%%%%%%%%%%%%%%%%%%%%
We have extended the gravitational reheating framework to include a massive spin-2 dark matter candidate and quantified its production from both inflaton oscillations and the thermal bath. The helicity composition of the spin-2 final state plays a pivotal role in the production dynamics. In particular, we find that the longitudinal (helicity-0) mode overwhelmingly dominates the gravitational production rate in the regime of light spin-2 mass. Analytically, the helicity-0 contribution to the matrix element grows as $\tau^{-4}$ (with $\tau \equiv m_2^2/m_\phi^2$) for $\tau \ll 1$, whereas the vector-like (helicity $\pm1$) part scales as $\tau^{-2}$ and the tensor (helicity $\pm2$) part tends to $\mathcal{O}(1)$. This leads to a hierarchy of yields: as $m_2$ decreases relative to the inflaton mass $m_\phi$, the longitudinal mode becomes dominant, for example, even at $\tau \simeq 1/2$ (i.e. $m_2 \simeq m_\phi/\sqrt{2}$) the helicity-0 channel already accounts for roughly $2/3$ of the total production rate. For extremely small $m_2$, the spin-2 production is enhanced so dramatically that the perturbative Fierz-Pauli description would enter a strong-coupling regime. This behavior is in stark contrast to lower-spin dark matter scenarios: for instance, spin-3/2 (gravitino-like) production via gravity is enhanced by a factor $\sim m_\phi/m_{3/2}$, and lower-spin cases do not exhibit the mass term in the denominator. The spin-2 case therefore represents the most parametrically efficient gravitational portal, with its small $m_2$ production cross section growing faster than any lower-spin counterpart.

We have calculated the relic abundance of the spin-2 field from reheating-era production. The contribution from inflaton condensate annihilations is found to far exceed that from thermal plasma scatterings in all regions of parameter space we examined. The inflaton-sourced channel continues to dominate for generic reheating models, including scenarios with inflaton potentials $V(\phi) \propto \phi^k$ (for $k>2$) that yield an extended reheating period. Numerically, across plausible post-inflationary histories, we find that gravitational scatterings among inflatons produce an abundance of $X_{\mu\nu}$ sufficient to account for all of dark matter (for appropriate choices of $m_2$ and $T_{\rm RH}$), while the corresponding thermal production of $X_{\mu\nu}$ is typically negligible (often by many orders of magnitude). Only in extreme limits (e.g. exceedingly high $T_{\rm RH}$ approaching the Planck scale, or finely tuned cases where inflaton annihilation is kinematically suppressed) would the thermal channel become competitive, and such regimes lie outside the scope of our minimal scenario. 

We therefore conclude that inflaton-mediated gravitational interactions set the leading post-inflation production “floor” for massive spin-2 dark matter. Requiring that $X_{\mu\nu}$ not overclose the Universe yields constraints on the parameters. The enhancement from the helicity-0 mode means that for a given $m_2$, there is an upper bound on $T_{\rm RH}$ (or conversely, for a given $T_{\rm RH}$ there is a lower bound on $m_2$) beyond which $\Omega_{X}$ would exceed the observed dark matter density. For example, in typical scenarios with $m_\phi \sim 10^{13}$ GeV, a reheating temperature $T_{\rm RH} \sim 10^{12}$ GeV and a spin-2 mass in the vicinity of the inflaton mass (within a factor of a few) can yield $\Omega_{X} h^2 \sim 0.12$ naturally, whereas significantly lighter $m_2$ would overproduce dark matter unless $T_{\rm RH}$ is correspondingly lower. For stable $X_{\mu\nu}$, the lower mass bound given by $T_{\rm RH} \gtrsim T_{\rm BBN} \simeq 2~\mathrm{MeV}$, which implies that $m_2 \gtrsim 2.7 \times 10^{12}~\mathrm{GeV}$.

Overall, our results demonstrate that a massive spin-2 particle coupled solely through gravity can be produced in the correct abundance during reheating, serving as a viable dark matter candidate. The helicity-0 mode dynamics are the key distinguishing feature of this spin-2 case, leading to more restrictive cosmological constraints but also a robust production mechanism that is “guaranteed” given any inflationary reheating scenario. We have thereby placed spin-2 dark matter on equal footing with spin-0, 1, 1/2, and 3/2 cases in the context of gravitational portals, and identified the unique parametric dependencies (notably the $\tau$-scaling and dominance of longitudinal modes) that characterize the spin-2 reheating origin of dark matter.

\noindent {\bf Acknowledgments. }  The author would like to thank Marcos Garc{\' i}a, Edward Kolb, and Andrew Long for useful discussions. The work of S.V. was supported by the Kavli Institute for Cosmological Physics at the University of Chicago.

%%%%%%%%%%%%%%%%%%%%%%%%%%%%%%%%%%%%%%%%%%%%%%%%%
\section*{Appendix}
%%%%%%%%%%%%%%%%%%%%%%%%%%%%%%%%%%%%%%%%%%%%%%%%%
\appendix
\renewcommand{\thesubsection}{\Alph{subsection}}

%%%%%%%%%%%%%%%%%%%%%%%%%%%%%%%%%%%%%%%%%%%%%%%%%%%%%%
\section{Boltzmann equation for a decaying inflaton condensate}
\label{app:boltz}
%%%%%%%%%%%%%%%%%%%%%%%%%%%%%%%%%%%%%%%%%%%%%%%%%%%%%%
In this appendix, we derive the evolution equation for the energy density of an oscillating inflaton condensate that loses energy through perturbative decays and scatterings. In the main text, these
results are applied to the gravitational production of a massive spin-2 field, but the derivation itself is completely general and depends only on the kinematics of the inflaton background.

We assume that the inflaton condensate is spatially homogeneous and that its decay proceeds perturbatively. The phase space distribution of the condensate field, denoted by $\phi$, can then be written as
\begin{equation}
f_{\phi}(k,t) \; = \; (2\pi)^3 n_{\phi}(t)\,\delta^{(3)}(\mathbf{k}) \, ,
\end{equation}
where $n_{\phi}(t)$ is the instantaneous number density of the condensate quanta and the delta function reflects the fact that they are at rest in the comoving frame.

Neglecting Bose enhancement and Pauli blocking for the decay products of $\phi$, the integrated Boltzmann equation for the inflaton number density takes the form \cite{Nurmi:2015ema}
\begin{equation}
\label{eq:boltzm1}
\dot{n}_\phi+3 H n_\phi \; = \;
-\int d \Psi_{\phi, A, B}\,
  |\mathcal{M}|_{\phi \rightarrow A B}^2\,
  f_\phi(k, t)\, .
\end{equation}
Here $A$ and $B$ label the decay products,
$d \Psi_{\phi, A, B}$ is the phase space measure for the initial
condensate and the two final-state particles, and $\mathcal{M}$ is the
corresponding transition amplitude. In this setup we ignore
backreaction into the condensate, that is, we neglect processes that
repopulate the zero-mode of $\phi$.

The right-hand side of Eq.~\eqref{eq:boltzm1} can be expressed more
explicitly as
\begin{equation}
\begin{aligned}
d \Psi_{\phi, A, B}|\mathcal{M}|_{\phi \rightarrow A B}^2
&=
\sum_{n=1}^{\infty}
\frac{d^3 \boldsymbol{k}}{(2 \pi)^3 n_\phi(t)}
\frac{d^3 \boldsymbol{p}_A}{(2 \pi)^3 2 p_A^0}
\frac{d^3 \boldsymbol{p}_B}{(2 \pi)^3 2 p_B^0}\\
&\times (2 \pi)^4
\delta^{(4)}\!\left(p_n-p_A-p_B\right)\,
\left|\mathcal{M}_n\right|^2 \, ,
\end{aligned}
\end{equation}
where $\mathcal{M}_n$ is the amplitude associated with the $n$-th
oscillation mode of the inflaton field during a single period, and
describes the transition from the coherent condensate state
$\ket{\phi}$ to a two-particle final state $\ket{A,B}$.

For the inflaton condensate, we impose the normalization
\begin{equation}
\int \frac{d^3 \boldsymbol{k}}{(2 \pi)^3 n_\phi}\,
f_\phi(k, t)
\; = \; 1 \, .
\end{equation}
Using this relation and carrying out the integration
over $\mathbf{k}$ in Eq.~\eqref{eq:boltzm1} gives
\begin{equation}
\begin{aligned}
\dot{n}_\phi+3 H n_\phi
& =
-\sum_{n=1}^{\infty}
\int \frac{d^3 \boldsymbol{p}_A}{(2 \pi)^3 2 p_A^0}
      \frac{d^3 \boldsymbol{p}_B}{(2 \pi)^3 2 p_B^0} \\
& \times (2 \pi)^4
      \delta^{(4)}\!\left(p_n-p_A-p_B\right)\,
      \left|\mathcal{M}_n\right|^2,
\end{aligned}     
\end{equation}
where $p_n = (E_n, \mathbf{0})$ and $E_n$ is the effective energy
carried by the $n$-th mode of the oscillating condensate.

It is often more convenient to phrase the result in terms of the
energy density of the inflaton condensate. Introducing the inflaton
equation of state parameter $w_\phi = p_{\phi}/\rho_{\phi}$, the
evolution of $\rho_\phi$ can be written as
\begin{equation}
\frac{d\rho_{\phi}}{dt} + 3H(1+w_{\phi})\rho_{\phi}
\;=\;
-(1+w_{\phi})\,\Gamma_{\phi}\,\rho_{\phi}\,.
\end{equation}
The term on the right-hand side represents the energy transferred per
unit time and per unit physical volume from the condensate to its
decay products. We define
\begin{equation}
(1+w_\phi)\Gamma_\phi\rho_\phi
\;\equiv\;
\frac{\Delta E}{{\rm Vol}_4} \,,
\end{equation}
where ${\rm Vol}_4$ denotes a four-volume in spacetime and
\begin{equation}
\begin{aligned}
\Delta E 
\; \equiv \;
&\int\frac{d^3p_A}{(2\pi)^3 2p_A^0}
    \frac{d^3p_B}{(2\pi)^3 2p_B^0}
    (p_A^0+p_B^0)\\
&\times \left|
  \frac{1}{n!}
  \biggl<
   {\rm f}\Big|
   \big(i\int d^4x_1{\cal L}_{\rm int}\big)\cdots
   \big(i\int d^4x_n{\cal L}_{\rm int}\big)
   \Big|0
  \biggr>
\right|^2 \, .
\label{eq:DeltaE}
\end{aligned}
\end{equation}
Here ${\cal L}_{\rm int}$ is the interaction Lagrangian that mediates
the inflaton decay, and the final state $\ket{\rm f}$ consists of the
particles produced in the decay.

The energy transfer per unit spacetime volume can be reorganized as
\begin{equation}
\begin{aligned}
&\frac{\Delta E}{{\rm Vol}_4}
=
\int\frac{d^3p_A}{(2\pi)^3 2p_A^0}
    \frac{d^3p_B}{(2\pi)^3 2p_B^0}
    (p_A^0+p_B^0)\\
& \times \sum_{m_1+ \ldots + m_n >0}^{\infty}
  |{\cal M}_{m_1, \ldots ,m_n}|^2\,
  (2\pi)^4\,
  \delta^4 \!\left(\sum_i p_{\phi, m_i}-p_A-p_B \right)\, .
\end{aligned}
\end{equation}

From this expression one can read off an effective decay or scattering
rate for the condensate \cite{Kainulainen:2016vzv, Ichikawa:2008ne, GKMO2},
\begin{equation}
\begin{aligned}
\label{eq:decgeneral}
\Gamma_\phi
&=
\frac{1}{8 \pi\left(1+w_\phi\right) \rho_\phi}\,
\frac{1}{\mathcal{S}!} \sum_{m_1+ \ldots + m_n >0}^{\infty} |{\cal M}_{m_1, \ldots ,m_n}|^2\\
& \times
\left( E_{m_1}+ \ldots +E_{m_n} \right)\,
\beta_{m_1, \ldots ,m_n}(m_A, m_B) \, ,
\end{aligned}
\end{equation}
with the usual two-body phase space factor
\begin{equation}
\begin{aligned}
&\beta_{m_1, \ldots ,m_n}(m_A, m_B) 
\equiv \\
&\sqrt{\left(1-\frac{\left(m_A+m_B\right)^2}{(\sum_i E_{m_i})^2}\right)
      \left(1-\frac{\left(m_A-m_B\right)^2}{(\sum_i E_{m_i})^2}\right)} \, .
 \end{aligned}     
\end{equation}
The factor $\mathcal{S}$ accounts for identical particles in
the final state. In the main text, $A$ and $B$ are often taken to be
two identical spin-2 particles, which implies $\mathcal{S}=2$.

To make connection with the inflaton dynamics used in the paper, we decompose the oscillating inflaton field as
\begin{equation}
    \phi(t)
    \; \simeq \;
    \phi_0(t)\, \cdot \mathcal{P}(t) \, ,
    \label{Eq:oscillation}
\end{equation}
where $\mathcal{P}(t)$ captures the rapid oscillations and $\phi_0(t)$ is a slowly-varying envelope that redshifts with the expansion. Over a single oscillation period $\phi_0$ can be treated as approximately constant. The oscillatory piece is expanded in Fourier modes as
\begin{equation}
\mathcal{P}(t)
\; = \;
\sum_{n=-\infty}^{\infty}\mathcal{P}_n\,e^{-in\omega t} \, ,
\label{PFexpand}
\end{equation}
with $\omega$ the fundamental oscillation frequency of the inflaton.

For a potential of the form $V(\phi) \propto \phi^{k}$, one can show
that the inflaton behaves as a fluid with effective equation-of-state
parameter \cite{GKMO2, Turner:1983he}
\begin{equation}
    \label{eq:eominf}
    w_{\phi}
    \; = \;
    \frac{k-2}{k+2} \, .
\end{equation}
The effective inflaton mass during oscillations about the minimum is
\begin{equation}
\label{eq:infeffmass}
m_\phi^2(t)
\equiv
V^{\prime \prime}\left(\phi_0(t)\right)
=
k(k-1)\,\lambda\,M_P^2
\left(\frac{\phi_0(t)}{M_P}\right)^{k-2} \, ,
\end{equation}
and the inflaton energy density can be written as
\begin{equation}
\begin{aligned}
    \label{eq:infenden}
     \rho_{\phi}(t)
     & \; = \;
    V(\phi_0(t)) 
    \; = \; 
     \lambda M_P^4
    \left(\frac{\phi_0(t)}{M_P}\right)^{k}\\
     & \; = \;
    \frac{1}{k(k-1)}\,m_{\phi}^2(t)\,\phi_0^2 \, .
\end{aligned}    
\end{equation}
These relations provide the bridge between the microscopic amplitudes
computed from the interaction Lagrangian ${\cal L}_{\rm int}$ and the
macroscopic evolution of the inflaton energy density. In the main
text we use this framework with ${\cal L}_{\rm int}$ given by the
universal gravitational couplings to the Fierz-Pauli stress-energy
tensor in order to compute the production rate of the massive spin-2
field for different choices of the inflaton potential and for various
values of $k$.

%%%%%%%%%%%%%%%%%%%%%%%%%%%%%%%%%%%%%%%%%%%%%%%%%%%%%%%%%%
\section{Thermal production of massive spin-2 particles}
\label{ap:thermal_spin2}
%%%%%%%%%%%%%%%%%%%%%%%%%%%%%%%%%%%%%%%%%%%%%%%%%%%%%%%%%%
In this appendix we derive the thermal production rate $R_{2}^T$ of the massive spin-2 field $X_{\mu\nu}$ from scatterings of relativistic Standard Model particles in the thermal bath. We restrict to the regime where all SM species in the initial state can be treated as massless and consider $2\to 2$ processes of the form
\begin{equation}
  \mathrm{SM} + \mathrm{SM} \;\rightarrow\; X + X \, .
\end{equation}
The corresponding dark matter production rate is given by the general
expression in Eq.~\eqref{thrate}, where we assume
$4m_{2}^{2} \ll s$ and include an overall factor of $2$ to account for the two spin-2 particles produced per scattering event.

The squared amplitudes are conveniently expressed in terms of the
Mandelstam invariants $s$ and $t$.  For the process $1+2\to 3+4$ in the centre-of-mass frame we take
\begin{equation}
  t \; = \; \frac{s}{2}
  \left(
    \sqrt{1 - \frac{4m_{2}^2}{s}} \cos{\theta_{13}} - 1
  \right)
  + m_{2}^2 \, ,
\end{equation}
\begin{equation}
  s \; = \; 2E_1 E_2\bigl(1- \cos{\theta_{12}}\bigr) \, ,
\end{equation}
where $\theta_{12}$ is the angle between the initial three-momenta and
$\theta_{13}$ is the scattering angle between $\mathbf p_1$ and
$\mathbf p_3$.

The gravity-mediated squared amplitudes for the SM initial states
enter through the averaged quantity in Eq.~\eqref{Eq:ampscat}.  We
include $4$ real degrees of freedom for the Higgs doublet, $12$
degrees for the gauge bosons (8 gluons plus 4 electroweak vectors) and $45$ degrees for the fermions (6 quark flavors with 3 colors and their antiparticles, 3 charged leptons and 3 neutrinos). The
expressions below are summed over final state polarizations and
include the symmetry factors for both initial and final states. This
is indicated by the bar on $|\overline{\mathcal{M}}|^2$.

For scalar initial states, the total squared amplitude for
gravity-mediated production of a spin-2 pair is
\begin{equation}
\begin{aligned}
&|\mathcal{\overline{M}}^{0 2}|^2 \; = \; \frac{1}{144 m_2^8 M_P^4 s^2} \Big[ \left(m_2^2-t\right)^2 \left(m_2^2-s-t\right)^2\times\\
& \times \left(180 m_2^8-120 m_2^6 s+46 m_2^4 s^2-8 m_2^2 s^3+s^4\right)  \Big]\, ,
\label{fullM2}
\end{aligned}
\end{equation}
where $m_2$ is the mass of the spin-2 particle.

For massless fermions in the initial state, the corresponding squared
amplitude reads
\begin{equation}
\begin{aligned}
&|\mathcal{\overline{M}}^{\frac{1}{2} 2}|^2 = \frac{1}{576 m_2^8 M_P^4 s^2} \times \Big[
\left(2 m_2^2-s-2 t\right)^2 \times \\
&\left(180 m_2^8-120 m_2^6 s+46 m_2^4 s^2-8 m_2^2 s^3+s^4\right) \\
& \left(2 m_2^2 t-m_2^4-t (s+t)\right) \Big]
\, .
\label{M1232}
 \end{aligned}    
\end{equation}
For initial massless gauge bosons, we obtain
\begin{equation}
\begin{aligned}
&|\mathcal{\overline{M}}^{1 2}|^2 = \frac{1}{72 m_2^8 M_P^4 s^2} \\
& \left(m_2^4-2 m_2^2 t+t (s+t)\right)^2 \\
&\left(180 m_2^8-120 m_2^6 s+46 m_2^4 s^2-8 m_2^2 s^3+s^4\right) 
\, .
\label{M132}
 \end{aligned}    
\end{equation}

Substituting Eqs.~\eqref{fullM2}-\eqref{M132} into the general expression for the thermal rate,
Eq.~\eqref{thrate}, and performing the phase-space integrals over the
initial-state momenta and scattering angle, one obtains a compact
high-temperature expansion for the total production rate of spin-2
particles,
\begin{equation}
\begin{aligned}
  \label{eq:rate2}
  R^{T}_{2}(T)
  &\; =\;
  \beta_{1}\,\frac{T^{16}}{m_{2}^{8} M_{P}^{4}}
  +\beta_{2}\,\frac{T^{14}}{m_{2}^{6} M_{P}^{4}}
  +\beta_{3}\,\frac{T^{12}}{m_{2}^{4} M_{P}^{4}}
  \\
  & \quad
  + \beta_{4}\,\frac{T^{10}}{m_{2}^{2} M_{P}^{4}}
  +\beta_{5}\,\frac{T^{8}}{M_{P}^{4}}
  +\beta_{6}\,\frac{m_{2}^{2}T^{6}}{M_{P}^{4}}
  \\
  & \quad
  + \beta_{7}\,\frac{m_{2}^{4}T^{4}}{M_{P}^{4}} \, ,
\end{aligned}
\end{equation}
where the dimensionless coefficients $\beta_i$ arise from the angular
integration and the thermal averaging over massless SM species:
\begin{equation}
  \beta_{1}
  \; = \; \frac{2560813\,\pi^{11}}{108864000} \, ,
\end{equation}
\begin{equation}
  \beta_{2}
  \; = \; -\frac{2981985}{16\,\pi^{5}}\zeta(7)^{2}\, ,
\end{equation}
\begin{equation}
  \beta_{3}
  \; = \; \frac{1023611\,\pi^{7}}{76204800}\, ,
\end{equation}
\begin{equation}
  \beta_{4}
  \; = \; -\frac{2710889}{3840\,\pi^{5}}\zeta(5)^{2} \, ,
\end{equation}
\begin{equation}
  \beta_{5}
  \; = \;\frac{1323659\,\pi^{3}}{248832000} \,,
\end{equation}
\begin{equation}
  \beta_{6}
  \; = \;-\frac{1531}{512\,\pi^{5}}\zeta(3)^{2} \, ,
\end{equation}
\begin{equation}
  \beta_{7}
  \; = \; \frac{151}{18432\,\pi} \, .
\end{equation}
The leading $T^{16}/m_2^8 M_P^4$ term reflects the strong enhancement
of the cross section at high centre-of-mass energies, while subleading
terms encode successively higher powers of $m_2^2/s$ in the exact
result. Equation~\eqref{eq:rate2} provides the thermal production
rate entering the Boltzmann equation for the spin-2 abundance in the
radiation-dominated epoch.

\addcontentsline{toc}{section}{References}
\bibliographystyle{utphys}
\bibliography{references}

\end{document}